\documentclass[%
jmp,
 sd,%
 amsmath,amssymb,
 floatfix,
 reprint,%
]{revtex4-1}

\usepackage{dcolumn}
\usepackage{bm}
\usepackage{amsfonts}
\usepackage{graphicx}
\usepackage{color}
\usepackage{mathtools}

\newcommand{\dual}{\overline}

\newcommand{\bor}{\mathop{\mathord{\lor}\!\!\!\raise4pt\hbox{$\scriptscriptstyle 2$}\,}}
\newcommand{\band}{\mathop{\mathord{\land}\!\!\!\lower2pt\hbox{$\scriptscriptstyle 2$}\,}}

\newcommand{\inc}{\, || \,}

\newcommand{\CHAIN}[1]{\mathbf{#1}}

\newcommand{\INT}[2]{[\![\CHAIN{#1}, \CHAIN{#2}]\!]}
\newcommand{\subspace}[1]{\langle{#1}\rangle}
\newcommand{\A}{{\CHAIN{A}}}
\newcommand{\B}{{\CHAIN{B}}}
\newcommand{\C}{{\CHAIN{C}}}
\newcommand{\D}{{\CHAIN{D}}}
\newcommand{\OO}{{\CHAIN{O}}}
\renewcommand{\P}{{\CHAIN{P}}}
\newcommand{\Q}{{\CHAIN{Q}}}
\newcommand{\R}{{\CHAIN{R}}}
\renewcommand{\S}{{\CHAIN{S}}}
\newcommand{\T}{{\CHAIN{T}}}
\newcommand{\X}{{\CHAIN{X}}}
\newcommand{\Y}{{\CHAIN{Y}}}
\newcommand{\I}{{\CHAIN{I}}}
\newcommand{\PQ}{{\CHAIN{PQ}}}
\newcommand{\RS}{{\CHAIN{RS}}}

\newcommand{\red}{\color{black}}

\newcommand{\black}{\color{black}}

\DeclareRobustCommand{\concatbase}{\mathbin{\rotatebox[origin=c]{90}{\scalebox{.7}{(\kern1ex)}}}}
\DeclareRobustCommand{\concat}{\,\concatbase\,}
\newcommand{\concatplus}{\oplus}

\newcommand{\setunion}{{\,\cup\,}}
\newcommand{\setunionplus}{{\,\uplus\,}}

\newenvironment{definition}[1][Definition]{\begin{trivlist}
\item[\hskip \labelsep {\bfseries #1}]}{\end{trivlist}}

\newcommand{\qed}{\nobreak \ifvmode \relax \else
      \ifdim\lastskip<1.5em \hskip-\lastskip
      \hskip1.5em plus0em minus0.5em \fi \nobreak
      \vrule height0.75em width0.5em depth0.25em\fi}

\begin{document}

\preprint{AIP/123-QED}

\title[A Potential Foundation for Emergent Space-Time]{A Potential Foundation for Emergent Space-Time}
\thanks{This work was supported, in part,
by a grant from the
John Templeton Foundation.}

\author{Kevin H. Knuth}
 \altaffiliation[Also at ]{Department of Informatics, University at Albany (SUNY), Albany NY 12222, USA}
 \email{kknuth@albany.edu.}
\author{Newshaw Bahreyni}
\affiliation{
Department of Physics, University at Albany (SUNY), Albany NY 12222, USA
}%

\date{\today}

\begin{abstract}
\red
We present a novel derivation of both the Minkowski metric and Lorentz transformations from the consistent quantification of a causally-ordered set of events with respect to an embedded observer. Unlike past derivations, which have relied on assumptions such as the existence of a 4-dimensional manifold, symmetries of space-time, or the constant speed of light, we demonstrate that these now familiar mathematics can be derived as the unique means to consistently quantify a network of events.  This suggests that space-time need not be physical, but instead the mathematics of space and time emerges as the unique way in which an observer can consistently quantify events and their relationships to one another.  The result is a potential foundation for emergent space-time.
\black
\end{abstract}



\maketitle

\section{Introduction}
\red
The unification of space and time into the concept of space-time is the centerpiece of one of the greatest scientific revolutions of the last century \cite{Einstein:1905}\cite{Minkowski:1909}.  The relationship between space and time as demonstrated by Einstein relied on two postulates: the principle of relativity and the constancy of the speed of light \cite{Einstein:1905}.  The principle of relativity represents a requirement of consistency, whereas the constancy of the speed of light represents an experimentally-determined result.  One might wonder if it is at all possible to demonstrate the precise relationship between space and time through logical means by requiring consistency in the spirit of Einstein along with some other principle or principles more fundamental than something that requires experimental observation, such as the constant speed of light.  A serious effort in this direction was undertaken by Robb \cite{Robb:1936}.  However, since space-time has come to be thought of as something that physically exists, it would not make any sense from this view to derive its properties through purely logical means.  One would expect to need to know something about space-time to recover its properties.

However, more recently, the idea that space-time is neither physical nor fundamental has been growing \cite{Seiberg:2006}.  The idea is that space and time may emerge from more fundamental relations or phenomena.  In one sense, this idea is not necessarily new. In addition to the older ideas, such as \emph{space as a container} or \emph{space as a substance}, which have mostly dominated our perspectives of space, is the view that space represents a relation between objects.  This relational viewpoint, \emph{space as relation}, which was proposed by the 11th century Islamic philosopher Al-Ghaz\={a}l\={i}, by the 17th century philosopher and mathematician Leibniz, by Kant in the 18th century, and Poincare’ in the 19th century has captured relatively little interest since Newton proposed the concepts of absolute space and time, which proved to have great predictive success.

We demonstrate that concepts of space and time, and their precise relation to one another, can emerge as a representation of relations among causally-related events.  While we take causality as a postulate, we have demonstrated in other work \cite{Knuth:FQXI2013}\cite{Knuth:Info-Based:2014} that it is of benefit to push back further and consider the idea that directed particle-particle interactions enable one to define a causal ordering among related events.  The basic idea is that everything that is detected or measured is the direct result of something influencing something else.
We focus on an intentionally simplistic, but fundamental, picture of influence where we consider the process of influence to connect and order the \emph{act of influencing} and the \emph{act of being influenced}.  We refer to each of these two acts with the generic term \emph{event}, so that the event associated with the act of influencing \emph{causes} the event associated with the act of being influenced.  This generic process of influence, along with the notion of transitivity of such influence, allows events to be partially-ordered.  This results in a mathematical structure referred to as a \emph{partially-ordered set}, or a \emph{poset} for short, which is the same as a directed acyclic graph (DAG).  This mathematical structure simply \emph{encodes} the causal relationships among events.  In this exploration, we will not be concerned with differentiating between distinct types of influence and their corresponding events.  That is, we consider only one type of influence.  Moreover, events can be coarse-grained so that connectivity of the poset can change depending on the magnification level.  In this paper, the events discussed typically represent coarse-grained events as opposed to the fundamental microscopic events considered in our other work \cite{Knuth:FQXI2013}\cite{Knuth:Info-Based:2014}, which has been shown to result in Fermion physics and the Dirac equation.
\black

Partially-ordered sets of events ordered by causal influence were introduced by Bombelli et al. \cite{Bombelli-etal-causal-set:1987} and called \emph{causal sets} or \emph{causets}.  Over the last twenty years, causal sets have been championed by Sorkin \cite{Sorkin:2003}\cite{Sorkin:2006} and employed in approaches to quantum gravity.  As such, they are typically endowed with, or embedded within, a Minkowski geometry exhibiting Lorentz invariance \cite{Bombelli-etal-origin-lorentz:1989}.

\red
We approach the problem from another direction entirely.
We do not conceive of these events as having taken place in some kind of space or time.  Instead, we are focused on the concept that particles can interact with one another and these interactions define events as well as their causal order.  As a result the poset of events is taken to be the fundamental structure. Moreover, rather than endowing it with additional properties, our goal is simply to identify a consistent means by which events in the poset can be \emph{quantified},
where the process of quantification entails assigning numbers to elements, or sets of elements, of posets, with the aim of numerically representing their relationships to one another.  More precisely, the process of quantification consists of defining order-preserving maps from the poset elements to real numbers.
Whereas assumptions or hypotheses regarding the properties of events or relationships between events have the potential to be right or wrong, consistent quantification can only be useful or not useful.

From our previous studies in quantifying lattices, which are special cases of partially-ordered sets \footnote{Lattices are partially-ordered sets where every pair of elements has a unique least upper bound called a join and a greatest lower bound called a meet.  This enables one to consider the join and meet as algebraic operators so that all lattices are algebras.}, we have found that the lattice symmetries constrain quantification resulting in constraint equations representing familiar sum and product rules \cite{Knuth:laws}\cite{Knuth:duality}\cite{Knuth:WCCI06}\cite{Knuth:measuring}\cite{Knuth:infophysics}\cite{Knuth&Skilling:2012}, which in the past have either been postulated or derived by other means.  Most recently, we have demonstrated how the relations among quantum-mechanical experimental setups, which are described by two-terminal series parallel (TTSP) graphs, sufficiently constrain the pair-wise quantification of measurement sequences resulting in constraint equations, which amount to the complex sum and product rules \cite{GKS:PRA}\cite{GK:Symmetry} of the Feynman path integral formulation of quantum mechanics \cite{Feynman:1948}.
The relevant insight is that the symmetries exhibited by an order-theoretic structure impose constraints on any quantification scheme resulting in constraint equations that represent physical laws.

Here we consider a universe of causally-related events, mediated by interactions, in which are embedded multiple observers each represented by a chain of events.  We demonstrate, through detailed proofs, that consistent relationships among a set of observers restricts any attempt at consistent quantification to a class of quantification schemes where observer-observer consistency results in constraint equations that represent a discrete version of the Minkowski metric and Lorentz transformations.  
\black

In the next section we introduce the concept of a distinguished chain, which can be used to represent an observer or an embedded agent.  Quantification of events and closed intervals along chains is established in Section \ref{sec:chain-quantification}.  In Section \ref{sec:poset-quantification}, these results are extended to quantify poset events.  In Section \ref{sec:chain-induced-structure}, we discuss chain-induced structure in the poset and follow this with Section \ref{sec:quantification-of-generalized-intervals} where we address quantification of intervals between events.  There we demonstrate that consistent quantification with respect to multiple chains exhibiting a constant relationship with one another results in a metric analogous to the Minkowski metric and that transformation of the quantification with respect to one pair of chains to quantification with respect to another pair of chains results in a pair-transform, which is analogous to the Bondi k-calculus \red formulation of Lorentz transformations \black \cite{Bondi:1980}.  In Section \ref{sec:spacetime-picture}, we demonstrate how this results in the mathematics of flat space-time.  Finally, in Section \ref{sec:dot-prod}, we develop the concept of subspace projection, which \red gives rise \black to the dot product. Collectively, these results suggest that the concept of space-time geometry emerges as the unique way for an embedded observer or agent to \red consistently quantify \black a partially-ordered set of events.

\section{Events, Chains and Observers} \label{sec:observers}

Influence is bounded by two events: the action of influencing and the reaction \red to \black being influenced.  As such, influence can be viewed as a binary ordering relation, which relates pairs of events.  That is, if event $x$ influences event $y$, we write $x \leq y$, and generically read `$y$ \emph{includes} $x$'.  \red The ordering relation is postulated to satisfy the following properties for elements $x, y,$ and $z$:
\begin{equation*}
\begin{array}{ll}
    \mbox{For all}~~x,~~x \leq x & \mbox{~~~~~(\emph{Reflexivity})}\\
    \mbox{If}~~x \leq y~~\mbox{and}~~y \leq x,~~\mbox{then}~~x = y &
    \mbox{~~~~~(\emph{Antisymmetry})}\\
    \mbox{If}~~x \leq y~~\mbox{and}~~y \leq z,~~\mbox{then}~~x \leq z &
    \mbox{~~~~~(\emph{Transitivity})}
\end{array}
\end{equation*}
Taken together, these postulates result in a \emph{partially-ordered set}, or \emph{poset}, of events.  Given any pair of events, it is not always true that one event is influenced by the other.  In this case, we say that the events are \emph{incomparable} and write $x \inc z$.  Last, we note that $x < y$ is interpreted as $x \leq y$ where $x \neq y$. \black
\black

Some posets, such as lattices, possess symmetries that give rise to algebraic structures that can be used to guide consistent quantification \cite{Knuth:laws}\cite{Knuth:measuring}\cite{Knuth&Skilling:2012}\cite{GKS:PRA}\cite{GK:Symmetry}.  However, this is not the case for posets in general, where often the only structure present is the partial ordering itself.  Given that our present goal is to discover what minimal structure is necessary to obtain useful constraints on the \red consistent quantification \black of events, we propose to introduce additional structure simply by \emph{distinguishing} a set of events, such as a finite chain, and quantifying a subset of the poset with respect to the distinguished set.

A chain is a subset of poset elements where for every element $x$ and $y$ in the chain we have that either $x \leq y$ or $y \leq x$, so that the elements comprising the chain are totally ordered.  In other words, a chain consists of a set of events which occur in succession.
\red It is the mental image of an observer with a clock taking note of the time at which events are observed that leads us to refer to the distinguished chain as an \emph{observer chain}.  However, we stress that we are not directly assuming a notion of time.  Instead, we have a notion of succession, which derives from the notion of causality since a set of binary comparisons, such as $x \leq y$ and $y \leq z$, enables one to define a chain of successive events $x < y < z$.  We will demonstrate that the ordered events along a chain \emph{define} time. \black

In the following sections, we focus on mathematics and introduce a consistent means by which one can quantify elements on a chain, as well as the intervals between elements on a chain.  By extending such quantification to a subset of the poset of events, we obtain a set of mathematical relations that enable a quantitative description of events in general.

\section{Quantifying a Chain} \label{sec:chain-quantification}
We begin by considering the consistent quantification of a finite chain of elements by assigning numbers to chain elements as well as intervals of elements along the chain.  \red This will be accomplished by defining an order-preserving map from elements of the chain to real numbers. \black

\subsection{Valuations}
We quantify elements of a chain $\P$ by defining a functional $v_{\P}$, called an \red \emph{monotonic valuation}, \black that takes each element $p$ of the chain $\P$ to a real number \red $v_{\P}(p)$ \black such that for every $p_x, p_y \in \P$ where $p_x < p_y$ we have that $v_{\P}(p_x) \leq v_{\P}(p_y)$.  The potential equality in the valuation is intentional as it allows for coarse graining where successive elements may be assigned the same number.  The valuation $v_{\P}$ takes the $N$ elements of the finite chain $\P$, $p_1, p_2, \ldots, p_N$, to a sequence of real numbers $v_{\P}(p_1) \leq v_{\P}(p_2) \leq \ldots \leq v_{\P}(p_N)$.  To simplify the notation, we overload the symbol that labels an element, such as $p$, by using it to also represent the valuation $v_{\P}(p)$ assigned to that element.  It will be understood from context whether the $p$ refers to the element or its real-valued valuation.


\subsection{Closed Intervals} \label{sec:closed-intervals}
Any pair of elements $p_i$ and $p_j$ on the chain $\P$ defines a unique set of elements called a \emph{closed interval}, denoted $[p_i, p_j]_\P$, such that
\begin{eqnarray}
[p_i, p_j]_\P & := & \{p \in \P | p_i \leq p \leq p_j \} \\
&  =& \{p_i, p_{i+1}, \ldots, p_{j-1}, p_j\}, \nonumber
\end{eqnarray}
where the subscript $\P$ indicates that the elements of the set belong to the chain $\P$.  Closed intervals can be quantified by defining an isotonic valuation $\phi$, which is a functional that takes a closed interval $I_\P$ to a real number $\phi(I_\P)$ such that $\phi(J_\P) \leq \phi(I_\P)$ if $J_\P \subseteq I_\P$.

Since general rules apply to specific cases, one can constrain the form of a general rule by considering special cases.  Consider joining two intervals that share a single element, which can be written as a set union
\begin{equation}
[p_i,p_j]_{\P} \setunion [p_j,p_k]_{\P} = [p_i, p_k]_{\P}.
\end{equation}
By writing $I = [p_i, p_k]_{\P}$, $J = [p_i, p_j]_{\P}$, $K = [p_j, p_k]_{\P}$, we can rewrite the expression above as $I = J \setunion K$.
Since the interval $I$ is related to both intervals $J$ and $K$, consistent quantification requires that the valuation quantifying $I$ must be some function of the valuations quantifying intervals $J$ and $K$.  This relation can be expressed as
\begin{equation} \label{eq:lengths-uplus}
\phi(I) = \phi(J) \setunionplus \phi(K)
\end{equation}
where $\setunionplus$ represents an unknown \red function, to be determined, \black that takes the valuations assigned to \red the \black closed intervals $J$ and $K$ to the valuation assigned to \red $I = J \setunion K$. \black

We can now consider joining a third closed interval, $L = [p_k, p_m]_{\P}$ sharing only the element $p_k$ with $K = [p_j,p_k]_{\P}$, and note that the joining process obeys associativity so that the order in which closed intervals are joined does not matter
\begin{equation}
(J \setunion K) \setunion L = J \setunion (K \setunion L).
\end{equation}
By applying (\ref{eq:lengths-uplus}), we find that the function $\setunionplus$ must also be associative
\begin{equation}
(\phi(J) \setunionplus \phi(K)) \setunionplus \phi(L) = \phi(J) \setunionplus (\phi(K) \setunionplus \phi(L)),
\end{equation}
which is made more apparent by writing $\alpha = \phi(J)$, $\beta = \phi(K)$ and $\gamma = \phi(L)$ so that
\begin{equation} \label{eq:associativity}
(\alpha \setunionplus \beta) \setunionplus \gamma = \alpha \setunionplus (\beta \setunionplus \gamma).
\end{equation}
Equation (\ref{eq:associativity}) is a functional equation for the function $\uplus$, which is known as the \emph{associativity equation} \cite{Aczel:FunctEqns}\cite{aczel+dhombres:1989}\cite{Knuth&Skilling:2012}.  Its general solution is
\begin{equation} \label{eq:associativity-soln}
\alpha \setunionplus \beta = f^{-1}(f(\alpha) + f(\beta))
\end{equation}
where $f$ is an arbitrary invertible function \cite{Aczel:FunctEqns}\cite{aczel+dhombres:1989}\cite{Knuth&Skilling:2012}. \red This implies that the function $\setunionplus$ in (\ref{eq:lengths-uplus}) is some invertible transform of additivity. \black  That is, there exists an invertible function $f$, which allows one to perform a regraduation of the valuation $\phi$ to a more convenient valuation $d = f \circ \phi$ so that $d(I) = f(\phi(I))$ and
\begin{equation} \label{eq:additivity-of-closed-intervals}
d(I) = d(J) + d(K),
\end{equation}
whenever $I = J \setunion K$ and $J \cap K$ is a singleton set.\footnote{While this result may seem obvious to some, it should be pointed out that additivity is a \emph{postulate} of measure theory.  Here we \emph{prove} that addition can be employed without a loss of generality, which is critical in establishing the resulting mathematics as a unique (up to invertible transform) means of quantifying intervals along a chain.}  We refer to this valuation $d$ of closed intervals along a chain as the \emph{length} of the interval.

We now consider closed intervals \red such as
\begin{equation}
[a,c]_{\P} = [a,b]_{\P} \setunion [b,c]_{\P}.
\end{equation}
Since the closed interval is defined by its endpoints, we require its length to be a function $s$ of the valuations assigned to the endpoints of the closed interval:
\begin{equation} \label{eq:length-and-endpoints}
d([a,b]_{\P}) = s(a,b),
\end{equation}
where the function $s$ is to be determined, the $a$ and $b$ in $d([a,b]_{\P})$ represent the chain elements, and the $a$ and $b$ in $s(a,b)$ represent the valuations of the chain elements.
By applying (\ref{eq:additivity-of-closed-intervals}) we have a functional equation for the function $s$
\begin{equation}
s(a,c) = s(a,b) + s(b,c),
\end{equation}
which has as its solution
\begin{equation} \label{eq:closed-interval-endpoint-difference}
s(a,c) = g(c)-g(a),
\end{equation}
where $g$ is an arbitrary function, since $s(a,c)$ cannot depend on $b$.

Since $g$ is an arbitrary function, we lose no generality by taking it to be the identity, so that $s(a,b) = b-a$.  The result is that \red we have proved that we lose no generality by taking the length of the interval to be the difference of the valuations assigned to its endpoints
\begin{eqnarray} \label{eq:length}
d([p_j, p_k]_\P) & = & v_{\P}(p_k) - v_{\P}(p_j) \\
& \equiv &  p_k - p_j, \nonumber
\end{eqnarray}
which is the usual lattice distance function \cite[p. 230]{Birkhoff:1967}.  \red It should be noted that we have not simply shown that one \emph{can} quantify intervals this way, we have shown that one \emph{must} quantify intervals this way in the sense that any quantification scheme that one could use must be an invertible transform of (\ref{eq:length}). \black

\section{Quantification of a Poset by Chain Projection} \label{sec:poset-quantification}
In this section, we focus on mathematics and examine the quantification of a partially-ordered set by introducing a method which we call \emph{chain projection}.  We limit ourselves to partially-ordered sets $\Pi$ in which a finite chain $\P$ comprised of $N$ distinct elements \red
\begin{equation}
p_1 < p_2 < \ldots < p_N \in \Pi
\end{equation} \black
can be identified \footnote{Here $p_i < p_j$ indicates that both $p_i \leq p_j$ and $p_i \neq p_j$.} \red and, using the results of Section \ref{sec:chain-quantification}, \black introduce a consistent method by which the chain can be quantified.  By introducing the notion of a projection onto a chain, we extend the quantification of a chain to a subset of the poset.  Note that we will find that it is not guaranteed that the entire poset can be quantified in such a manner.  \red While the result holds in general for all such posets, this fact has important consequences for the resulting physics of events and the mathematics of space and time. \black

\subsection{Chain Projection Mapping}
We consider the possible relationships between the chain $\P$ and elements of the poset $\Pi$.  Given an element $x \in \Pi$ and $x \notin \P$, one of the four following situations (illustrated in Figure \ref{fig:poset-chain-relations}) holds:

\begin{figure}[t]
  \begin{center}
  \includegraphics[height=0.33\textheight]{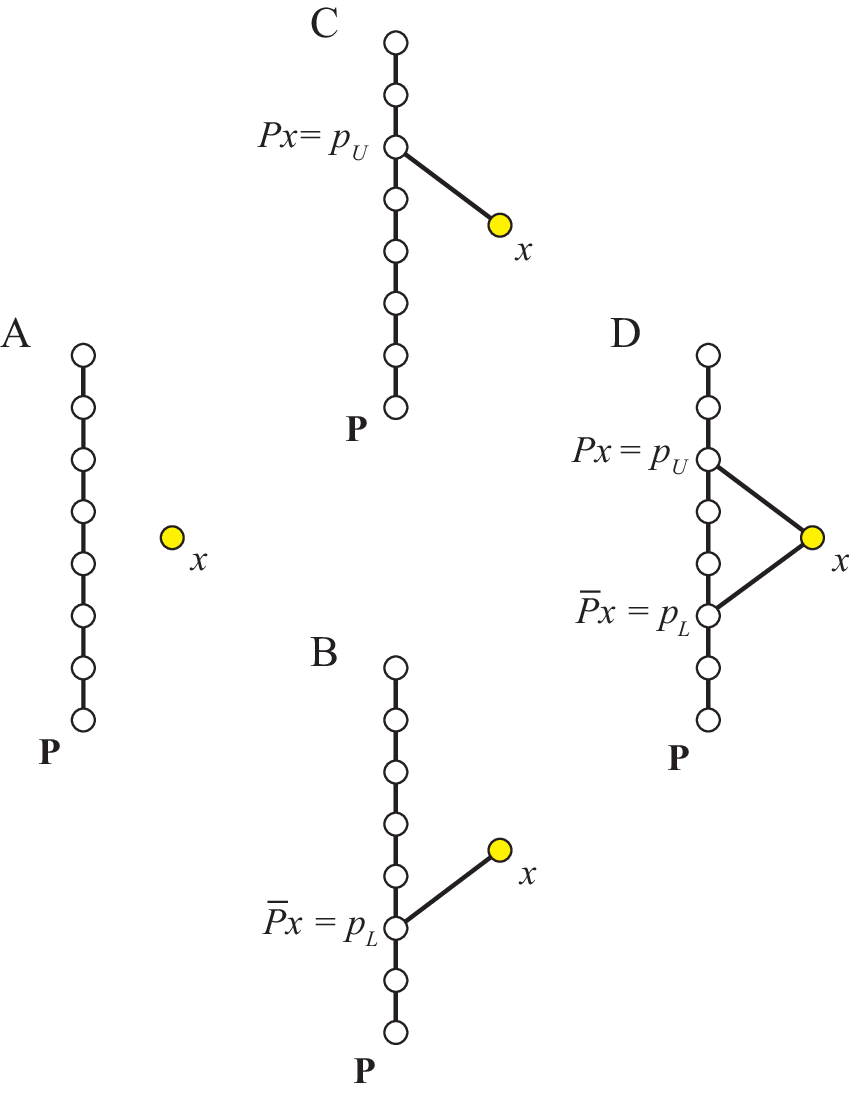}
  \end{center}
  \caption{Illustration of the four possible relationships between a chain $\P$ and an element $x \notin \P$.  Note that in (D) the element $x$ can be mapped to two elements on $\P$ using the projection operators $P$ and $\dual{P}$ so that $\dual{P}x = p_L$ and $Px = p_U$.  Poset elements exhibiting this relationship to the chain can be quantified in two ways by inheriting the quantification $v_\P(\dual{p}_x)$ and $v_\P(p_x)$ of the elements $\dual{p}_x$ and $p_x$.}
  \label{fig:poset-chain-relations}
\end{figure}

\begin{equation} \nonumber
\begin{array}[t]{cl}
    A. & \mbox{$\P$ and $x$ are incomparable} \\
    &
        \left\{
            \begin{array}{lll}
                p_i \inc x \quad & \forall & \quad 1 \leq i \leq N
            \end{array}
        \right.
    \\
    \\
    B. & \mbox{$x$ includes elements of $\P$} \\
    &
         \left\{
            \begin{array}{lll}
                p_i \leq x \quad & \forall & \quad 1 \leq i \leq L \\
                p_i \inc x \quad & \forall & \quad L < i \leq N
            \end{array}
        \right.
    \\
    \\
    C. & \mbox{elements of $\P$ include $x$} \\
    &
        \left\{
            \begin{array}{lll}
                p_i \inc x \quad & \forall \quad & 1 \leq i < U \\
                p_i \geq x \quad & \forall \quad & U \leq i \leq N
            \end{array}
        \right.
    \\
    \\
    D. & \mbox{$\P$ and $x$ include one another} \\
    &
       \left\{
            \begin{array}{lll}
                p_i \leq x \quad & \forall & \quad 1 \leq i \leq L \\
                p_i \inc x \quad & \forall & \quad L < i < U \\
                p_i \geq x \quad & \forall & \quad U \leq i \leq N
            \end{array}
        \right.
\end{array}
\end{equation}

We say that the poset element $x$ can be \emph{projected} onto a chain $\P$ if there exists an element $p \in \P$ such that $x \leq p$ (Cases C and D).  If this is the case, then the \emph{forward projection}, or simply \emph{projection}, of $x$ onto the chain $\P$ is given by the least event $p_x$ on the chain $\P$ such that $x \leq p_x$ where $p_x := \min \{p \in \P | x \leq p\}$, which is indicated by the index $U$ above.  Since the projection of an element onto a chain, if it exists, is unique, we can define the projection to be a functional $P : x \in \Pi \rightarrow p_x \in \P$ where the domain is $\{x | x \leq p_{max} \}$ where $p_{max}$ is the greatest element of the chain.  By applying this functional to the element $x$, we have $p_x = Px$.  Note that the functional is named after the chain.  Similarly, one can define a projection map $Q$ onto \red another \black chain $\Q$.

One can also consider the \emph{dual projection} \footnote{Note that the dual projection $\dual{P}x$ of $x$ onto the chain $\P$ is equal to the projection of $x$ onto the dual chain $\P^\partial$ where the order is reversed.} or \emph{backward projection} $\dual{P}$ where one identifies the greatest element $\dual{p}_x$ on the chain that is included by the poset element $x$.  We can define a functional $\dual{P}$ that takes elements $x$ from the domain $\{x | p_{min} \leq x \}$, where $p_{min}$ is the least element of the chain, to an element of the chain given by $\dual{p}_x = \dual{P}x := \max \{p \in \P | p \leq x\}$. This corresponds to the element indexed by $L$ above (Cases B and D).

\begin{figure}[t]
  \begin{center}
  \includegraphics[height=0.25\textheight]{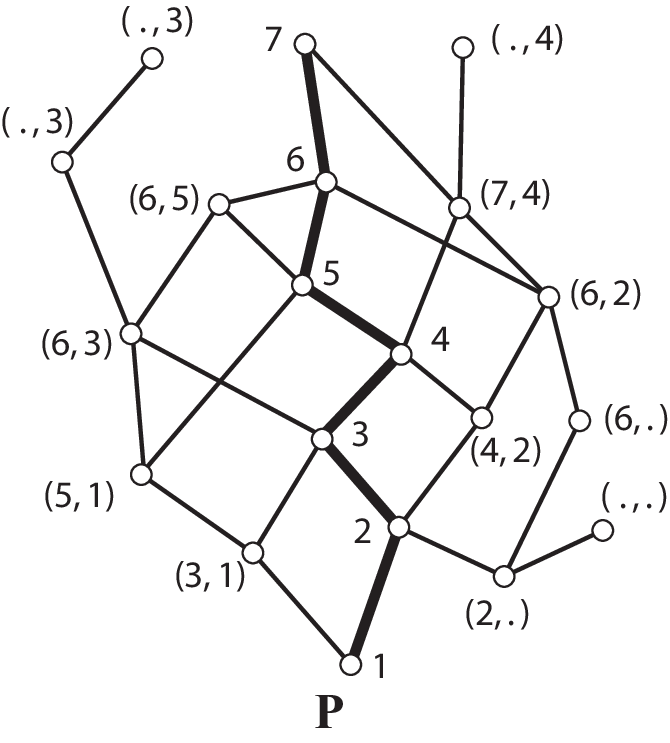}
  \end{center}
  \caption{This figure illustrates the quantification of a poset with respect to the chain $\P$.  The chain is quantified by an isotonic valuation of successive integers.  Poset elements, such as $x$, are then quantified by pairs of the form $(p_x, \dual{p}_x)$ representing the forward and backward projections onto the chain $\P$.  Note that not all elements can be quantified by two numbers.  Some elements, such as those quantified by $(.,3)$ do not forward project to the chain $\P$. Others, such as $(2,.)$ do not possess a backward projection.  Last, some elements, such as the one quantified by $(.,.)$ are incomparable to all elements of the chain and cannot be quantified \red by the chain $\P$ \black.}
  \label{fig:quantified-poset}
\end{figure}

\red
The forward and backward projections, $P$ and $\dual{P}$ take \black poset elements in their respective domains to elements on the chain $\P$. We can define a functional called the \emph{chain projection map} that takes a poset element $x$ in the intersection of the domains of $P$ and $\dual{P}$ and maps it to an ordered pair of chain elements (Case D) given by $(Px, \dual{P}x)$.  The chain projection map provides information about the connectivity of a poset from the perspective of the observer chain.  By composing the chain projection map with the valuation map from chain elements to real numbers, the valuation can be extended from the chain itself to poset elements that both forward and backward project onto the chain.  That is, a poset element $x$ that projects to $(Px, \dual{P}x)$ can be quantified by the pair of real numbers $(v_\P(Px), v_\P(\dual{P}x))$.  Figure \ref{fig:quantified-poset} illustrates quantification of a poset by chain projection.  Such quantification results in a set of \emph{chain-based coordinates} that are dependent on the particular chain used for quantification.  Note that since an element $x$ on the chain $\P$ satisfies $x = Px = \dual{P}x$, it is quantified by the pair $(v_\P(x), v_\P(x))$.  While the chain projection map does not ensure that all elements in the partially-ordered set will be quantified, we will find it to be extremely useful.

\subsection{Generalized Intervals}
We extend the concept of a closed interval on an ordered chain to that of a \emph{generalized interval}.  A generalized interval, denoted $[x,y]$, is identified by an ordered pair of elements $x,y \in \Pi$, each of which is called an \emph{endpoint}.  Note that $x$ and $y$ may be elements of different chains, and thus are either comparable or incomparable.  For this reason, the subscript referring to a chain has been dropped from the notation. Henceforth we shall refer to a generalized interval simply as an \emph{interval}.

The remainder of the paper will focus on deriving a quantification of intervals via chain projection.  However, before we consider such quantification, we first explore the structure revealed by distinguishing an observer chain.

\section{Chain-Induced Structure} \label{sec:chain-induced-structure}

Together, the act of distinguishing an observer chain and the technique of chain projection effectively map a subset of poset elements, and hence intervals, onto chain elements and closed intervals, respectively.  This reveals information about the connectivity of the poset of events from the perspective of the distinguished chain by specifically indicating the relationship between the chain and the poset elements in the quantified subset.  As \red such, by considering the poset along with the distinguished chain, \black the chain projection mapping has the effect of inducing structure in a poset \red which alone \black may lack any inherent characteristic structure or symmetry.

In this section, we explore the structure induced by such mappings in the case of multiple quantifying chains where closed intervals along one quantifying chain project to closed intervals along another quantifying chain.  For arbitrary chains in a general poset, one does not expect there to be a relationship between such projections.  However, in the special case where a set of multiple quantifying chains mutually agree in their quantification of one another we show that there exists a unique consistent pair-wise and scalar quantification of intervals.  This sets the stage for an observer-based geometry.

\subsection{Induced Subspaces} \label{sec:induced-subspaces}
In this section we demonstrate how multiple chains can induce a subspace within the poset.
We begin by introducing the concept of collinearity.

\begin{figure*}[t]
  \begin{center}
  \includegraphics[height=0.45\textheight]{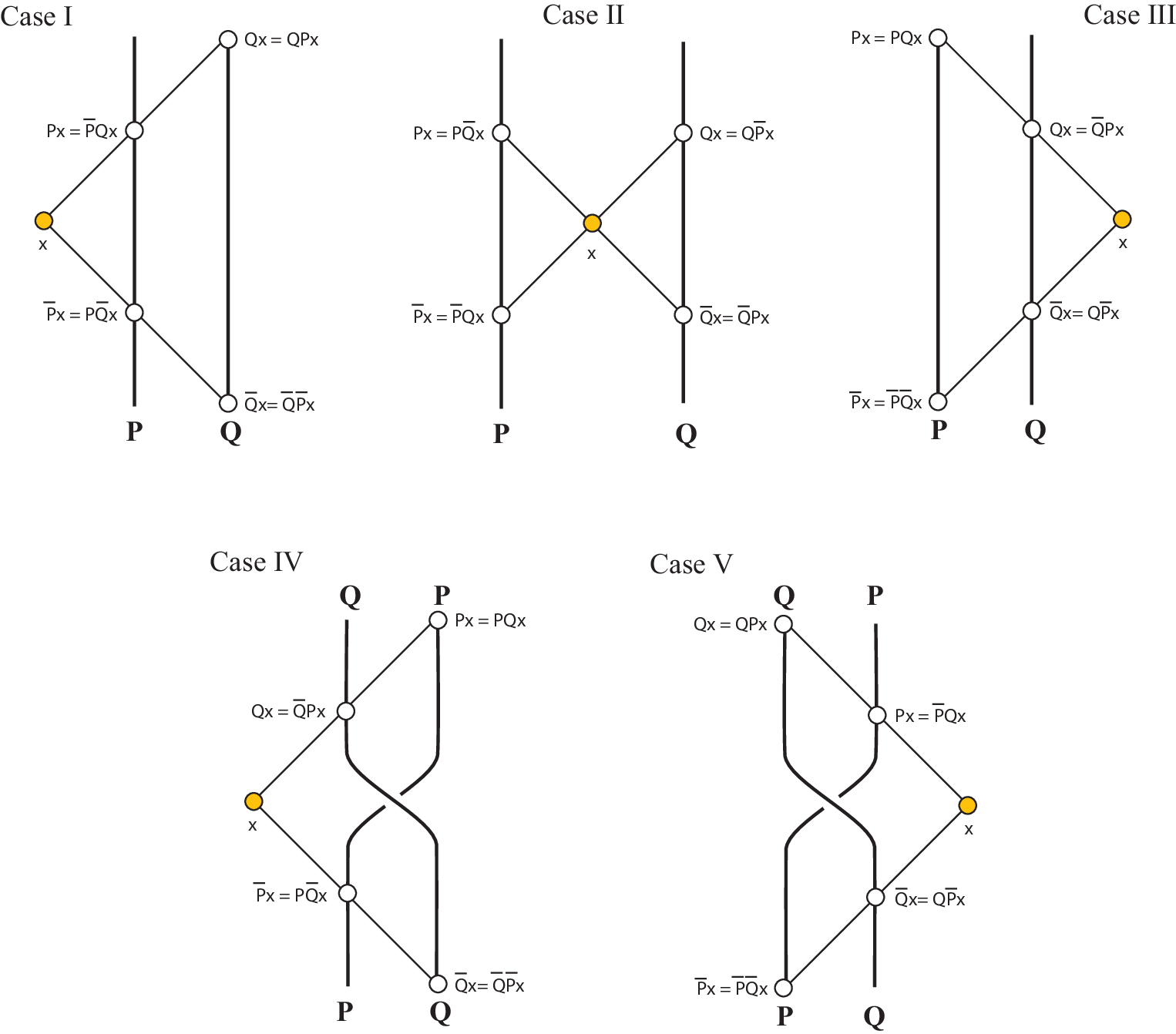}
  \end{center}
  \caption{An illustration of the five ways in which an element $x$ can be collinear with its projections onto two distinct finite chains.  In this way, an element can be said to be either on one side or the other of, or between, a pair of chains.  This introduces the concept of directionality to the subspace generated by the coordinated chains $\P$ and $\Q$.}
  \label{fig:collinearity}
\end{figure*}

\begin{definition}[Collinearity]
\red An element $x$ that possesses both backward and forward projections onto two distinct finite chains $\P$ and $\Q$ is said to be \emph{collinear} with those projections iff each of its projections onto $\P$ can be found by first projecting onto $\Q$ and then onto $\P$, and vice versa.\black
\end{definition}
There are five ways that this can be \red realized \black without violating the partial order (illustrated in Figure \ref{fig:collinearity}):
\begin{equation} \label{eq:collinearity_cases}
\begin{array}{c@{}c}
    \begin{array}{ccl}
        Px = \dual{P}Qx & \qquad Qx = QPx \\
        \dual{P}x = P\dual{Q}x & \qquad \dual{Q}x = \dual{Q} \, \dual{P}x
    \end{array} & \qquad \mbox{(Case I)} \\
    \\
    \begin{array}{ccl}
        Px = P\dual{Q}x & \qquad Qx = Q\dual{P}x \\
        \dual{P}x = \dual{P}Qx & \qquad \dual{Q}x = \dual{Q}Px
    \end{array} & \qquad \mbox{(Case II)} \\
    \\
    \begin{array}{ccl}
        Px = PQx & \qquad Qx = \dual{Q}Px \\
        \dual{P}x = \dual{P} \, \dual{Q}x & \qquad \dual{Q}x = Q\dual{P}x
    \end{array} & \qquad \mbox{(Case III)} \\
    \\
    \begin{array}{ccl}
        Px = PQx & \qquad Qx = \dual{Q}Px \\
        \dual{P}x = P\dual{Q}x & \qquad \dual{Q}x = \dual{Q} \, \dual{P}x
    \end{array} & \qquad \mbox{(Case IV)} \\
    \\
    \begin{array}{ccl}
        Px = \dual{P}Qx & \qquad Qx = QPx \\
        \dual{P}x = \dual{P} \, \dual{Q}x & \qquad \dual{Q}x = Q\dual{P}x
    \end{array} & \qquad \mbox{(Case V)} \\
\end{array}
\end{equation}
While these situations enable one to relate the element $x$ to its projections onto the chains, it is impossible to describe its relation to the finite portions of a chain between backward and forward projections of $x$.  Cases IV and V highlight one of the complications that arises where the chains perform a half-twist relative to $x$ in the segment between the projections.  One could imagine that something like this could happen in Cases I-III as well since the chains could undergo an integral number of full-twists in the segments between the forward and backward projections of $x$.  However, such behavior could only be defined by projecting elements other than $x$ onto the chains.  Considering $x$ alone, such possibilities are irrelevant.

Note that Cases I-III are invariant with respect to interchange of the forward and backward projections.  In other words the sub-poset defined by the chains and the element $x$ is dual to itself with respect to the ordering relation.  This leads us to the more refined concept of proper collinearity, which enables us to consistently order the chains with respect to the forward and backward projections of element $x$.
\begin{definition}[Proper Collinearity]
An element $x$ \red that possesses both backward and forward projections onto two distinct finite chains $\P$ and $\Q$ is said to be \emph{properly collinear} with its projections onto $\P$ and $\Q$ \black iff it is collinear with its projections onto the two chains and those projections are invariant with respect to reversing the ordering relation.
\end{definition}

We can extend the concept of proper collinearity to chains.
\begin{definition}[Proper Collinearity of Chains]
A finite chain $\X$, with least element $x_{min}$ and greatest element $x_{max}$, is said to be \emph{properly collinear} with $\P$ and $\Q$ iff each element $x \in \X$ is properly collinear with its projections onto $\P$ and $\Q$, and these projections constitute a surjective map from $\X$ onto the finite subchains defined by the closed intervals $[Px_{max}, \dual{P}x_{min}]_{\P}$ and $[Qx_{max}, \dual{Q}x_{min}]_{\Q}$.
\end{definition}
This leads to a geometric interpretation of the three relationships seen in Cases I-III above.  To begin to understand this, the concept of proper collinearity enables one to divide the poset into two equivalence classes based on whether or not an element is properly collinear with the chains $\P$ and $\Q$.  Elements that are properly collinear with $\P$ and $\Q$ are said to reside in the subspace defined by $\P$ and $\Q$, denoted $\subspace{\P\Q}$.  This subspace can be further divided into three equivalence classes based on Cases I-III.  An element $x$ exhibiting the relationship in Case I is said to be situated on the $\P$-side of its projections onto the pair of chains $\P$ and $\Q$, which is denoted by $x|\P|\Q$; whereas an element $x$ exhibiting the relationship in Case III is said to be situated on the $\Q$-side of of its projections onto the pair of chains $\P$ and $\Q$, which is denoted by $\P|\Q|x$.  Last, an element $x$ exhibiting the relationship in Case II is said to be situated \emph{between} its projections onto $\P$ and $\Q$, which is denoted by $\P|x|\Q$.

These relationships can be extended to any generalized interval $[a,b]$ resulting in nine cases, three cases where the two elements are situated similarly
\begin{equation}
\begin{array}{ll}
[a,b]|\P|\Q & \quad \mbox{$[a,b]$ is on the $\P$-side of $\P$ and $\Q$} \\
\P|\Q|[a,b] & \quad \mbox{$[a,b]$ is on the $\Q$-side of $\P$ and $\Q$} \\
\P|[a,b]|\Q & \quad \mbox{$[a,b]$ is between $\P$ and $\Q$}
\end{array}
\end{equation}
and six cases where the interval $[a,b]$ straddles one or more chains
\begin{equation}
\begin{array}{lr}
a|\P|b|\Q & b|\P|a|\Q\\
a|\P|\Q|b & b|\P|\Q|a\\
\P|a|\Q|b & \P|b|\Q|a.
\end{array}
\end{equation}

\begin{figure}[t]
  \begin{center}
  \includegraphics[height=0.20\textheight]{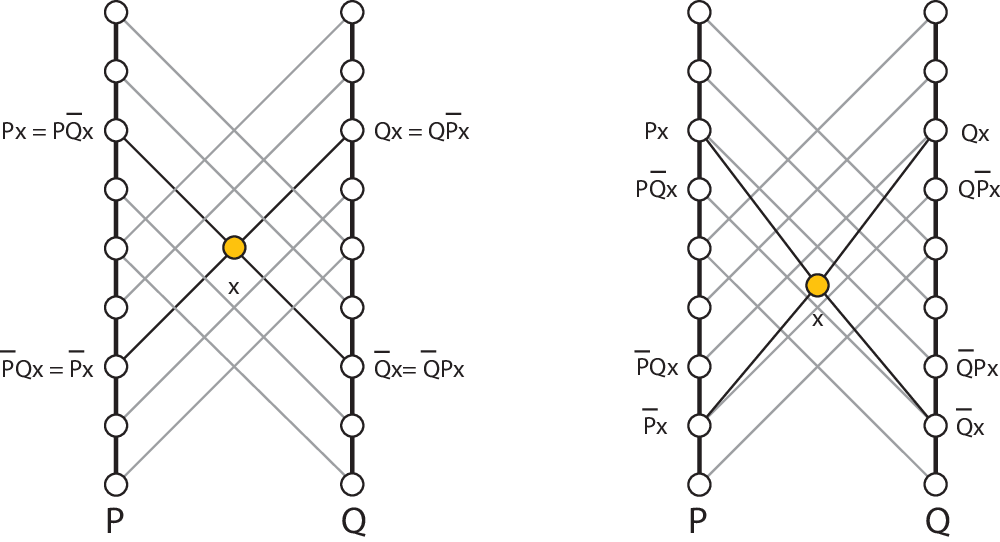}
  \end{center}
  \caption{(Left) The element $x$ is not only properly collinear with, but also between, chains $\P$ and $\Q$ since $Px = P\dual{Q}x$ and $Qx = Q\dual{P}x$. In this situation, quantification can be performed by considering only forward projections, such as $Px$ and $Qx$. (Right) An example of an element $x$ which is not collinear with the chains $\P$ and $\Q$.  In this sense, the chains $\P$ and $\Q$ define a subspace, which can include or exclude elements.  Note that the definition of the projection imposes the constraint that $P\dual{Q}x \leq Px$, and similarly for the other pairs of projections.}
  \label{fig:between}
\end{figure}

In the case of properly collinear finite chains, $\X$ can be said to be either on the $\P$-side of $\P$ and $\Q$ as in $\X|\P|\Q$, or between $\P$ and $\Q$ as in $\P|\X|\Q$, or on the $\Q$-side of $\P$ and $\Q$ as in $\P|\Q|\X$.  This introduces a way in which some chains belonging to this subspace can be ordered, which in turn induces bi-directionality. We now have an induced ordering relation where given any triple of distinct finite chains $\A, \B, \C$ obeying $\A|\B|\C$, we can write
\begin{equation}
\A|\B|\C \quad \Rightarrow \quad
\begin{array}{c}
    \A < \B < \C \\
    \mbox{or} \\
    \C < \B < \A
\end{array}
\end{equation}
where the directionality of the induced ordering relation is arbitrary, and the symbol $<$ in $A < B$ represents $A \leq B$ and $A \neq B$.  Note that in this case the binary ordering relation between pairs of chains is defined by their relation to a third chain.  Again, it is the introduction of an agent (the third chain) that enables one to define an ordering relation between two chains.  It is in this sense that we say that the structure is induced by the distinguished chain.

The act of identifying a distinguished chain results in two distinct ordering relations: one along chains, and one among properly collinear chains.  For this reason, we say that the induced subspace obtained by considering these two ordering relations is 1+1 dimensional.


\subsection{Coordinated Chains} \label{sec:coordinated-chains}

In this section, we consider posets that support a set of properly collinear observer chains, which can be used to quantify the poset via chain projection.  Furthermore, we consider chains that are \emph{coordinated} so that they project onto one another in a well-defined manner thus enabling us to extend the chain-based coordinate system induced by each chain to a coordinate system that is potentially more global.  This relationship is clarified by two definitions:

\begin{definition}[Compatibility]
Two chains $\P$ and $\Q$ are said to be \emph{compatible} over the intervals $[\dual{p}_{min}, \dual{p}_{max}]_\P$, $[p_{min}, p_{max}]_\P$ and $[\dual{q}_{min}, \dual{q}_{max}]_\Q$, $[q_{min}, q_{max}]_\Q$ iff the elements in $[\dual{p}_{min}, \dual{p}_{max}]_\P$ forward project to the elements in $[q_{min}, q_{max}]_\Q$ and the elements in $[p_{min}, p_{max}]_\P$ back project to the elements in $[\dual{q}_{min}, \dual{q}_{max}]_\Q$ in such a way that these projections are bijective.
\end{definition}

\begin{definition}[Coordinated]
Two chains $\P$ and $\Q$ are said to be \emph{coordinated} over the intervals $[\dual{p}_{min}, \dual{p}_{max}]_\P$, $[p_{min}, p_{max}]_\P$ and $[\dual{q}_{min}, \dual{q}_{max}]_\Q$, $[q_{min}, q_{max}]_\Q$ iff they are compatible over those intervals, and if the length of a closed interval on $\P$ is equal to the length of its image on $\Q$ and vice versa.
\end{definition}
Two coordinated chains quantify each others' intervals in an identical manner.  Since, quantification of generalized intervals in the poset is performed by projecting the generalized interval onto two closed intervals on a distinguished chain, the fact that two coordinated chains project to one another and agree on the quantification of each others' closed intervals implies that these two coordinated chains must agree on the quantification of any generalized interval that projects into the coordinated range.

\begin{figure}[t]
  \begin{center}
  \includegraphics[height=0.20\textheight]{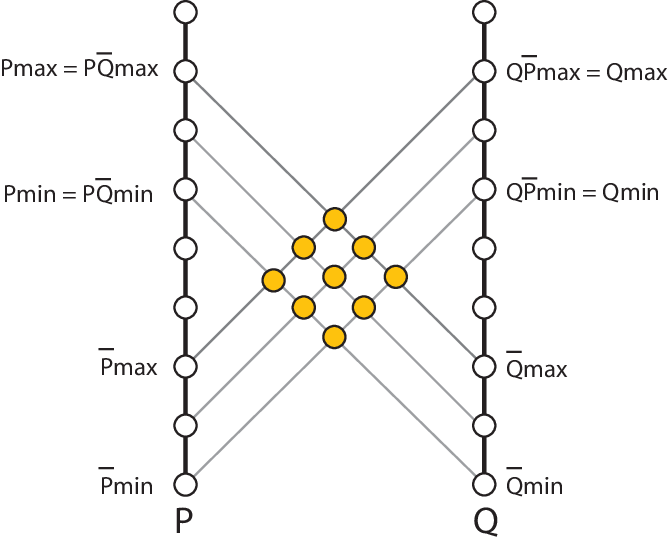}
  \end{center}
  \caption{An illustration of two chains coordinated over the range given by \red $[\dual{P}_{min}, \dual{P}_{max}]_\P$, $[P_{min}, P_{max}]_\P$ and $[\dual{Q}_{min}, \dual{Q}_{max}]_\Q$, $[Q_{min}, Q_{max}]_\Q$. \black Any interval formed from the indicated events in between the two chains can be quantified by the same pair whether it is obtained by forward and backward projections onto $\P$ or $\Q$, or by forward projections onto both $\P$ and $\Q$.}
  \label{fig:coordination-2}
\end{figure}

The concept of coordinated chains is illustrated in Fig. \ref{fig:coordination-2}.  A pair of coordinated chains $\P$ and $\Q$ generates a 1+1 dimensional subspace, denoted by $\subspace{\P\Q}$, which includes $\P$, $\Q$, and all elements collinear with $\P$ and $\Q$.  In this situation, the pair of chains can be used to quantify elements between them (Case II) using only forward projections, which is an advantage since in practical situations it represents information that can be obtained from the chains themselves.  This is because Case II of the collinearity condition requires that $Px = P\dual{Q}x$ and $Qx = Q\dual{P}x$ so that $v_\P(\dual{P}_y)-v_\P(\dual{P}_x) = v_\Q(Q_y)-v_\Q(Q_x)$ as required by the definition of coordinated chains.

\section{Quantification of Generalized Intervals} \label{sec:quantification-of-generalized-intervals}

\subsection{The Interval Pair}
In this section we consider the consistent quantification of intervals by a coordinated set of chains.  We will focus on situations where both elements defining the endpoints of the interval both forward project and back project onto each of the quantifying chains under consideration.  Since coordination is based on the projection of closed intervals situated on one chain to closed intervals on another chain, we begin by examining consistent quantification in that special case.

\begin{figure*}[t]
  \begin{center}
  \includegraphics[height=0.27\textheight]{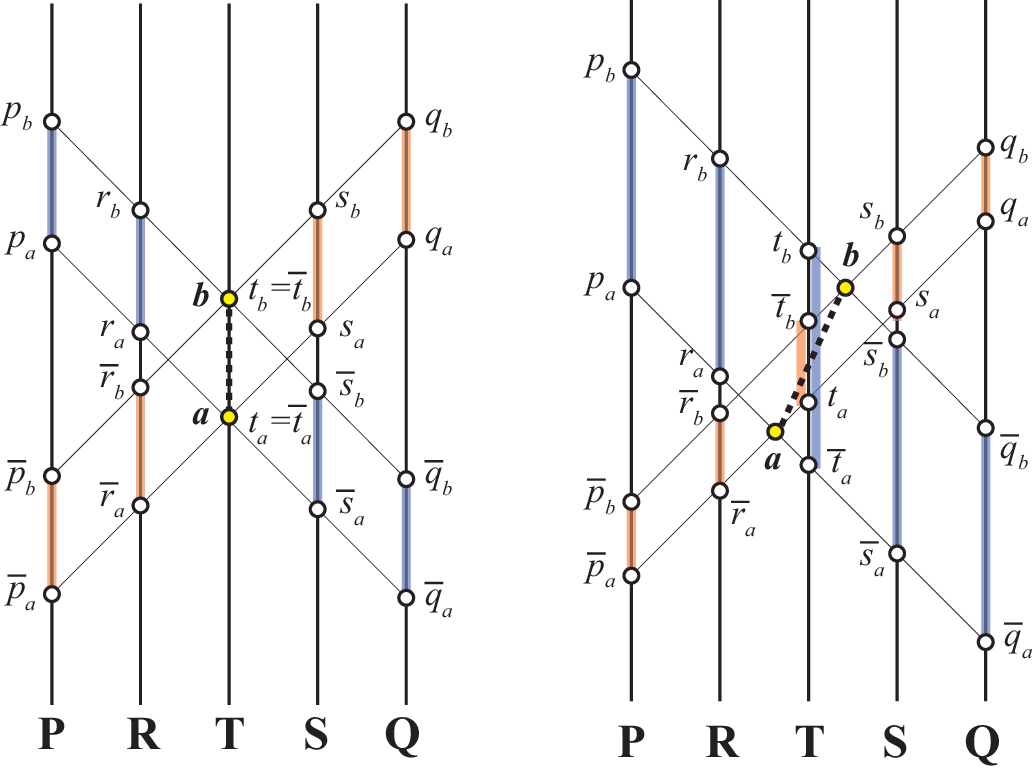}
  \end{center}
  \caption{(a) An illustration of the consistent quantification of a closed interval $[a,b]_\T$ on a chain that is a member of a coordinated set of chains.  The interval projects to a pair of intervals on each chain of the set.  This enables one to quantify the interval with either four values (4-tuple) or a pair of closed interval lengths, called an interval pair.  Note that, in this case, the interval pair consists of identical components.  As discussed in Section \ref{sec:IntervalClasses}, this closed interval is characterized as a pure chain-like interval by the other chains in the coordinated set. (b) A generalized interval also projects to a pair of intervals on each chain. However, its relationship to the quantifying chain determines the way in which the interval pair is consistently computed.  Because the components of the interval pair have like sign, this generalized interval is classified as a chain-like interval (see discussion in Section \ref{sec:IntervalClasses}).}
  \label{fig:pair-quantification-chainlike}
\end{figure*}

Figure \ref{fig:pair-quantification-chainlike} considers a set of five mutually coordinated chains: $\P$, $\Q$, $\R$, $\S$, and $\T$, which collectively form a 1+1 dimensional subspace.  These are labeled so that the chains $\R$, $\S$, and $\T$ are situated between the pair of chains $\P$ and $\Q$, and the chain $\T$ is situated between the pair of chains $\R$ and $\S$, so that $\P | \R | \T | \S | \Q$.  We consider two elements $a, b \in \T$ that form the closed interval $[a,b]_\T$.  This closed interval both forward projects and back projects onto each of the four other quantifying chains in the coordinated set.  By mapping the elements $a$ and $b$ onto the other chains, we can obtain a quantification of $[a,b]_\T$ based on the following \emph{4-tuples} of valuations:
\[
\begin{array}{llll}
(v_\P(Pa), & v_\P(Pb), & v_\P(\dual{P}a), & v_\P(\dual{P}b))_\P \\
(v_\R(Ra), & v_\R(Rb), & v_\R(\dual{R}a), & v_\R(\dual{R}b))_\R \\
(v_\T(a),  & v_\T(b),  & v_\T(a), & v_\T(b))_\T \\
(v_\S(Sa), & v_\S(Sb), & v_\S(\dual{S}a), & v_\S(\dual{S}b))_\S \\
(v_\Q(Qa), & v_\Q(Qb), & v_\Q(\dual{Q}a), & v_\Q(\dual{Q}b))_\Q,
\end{array}
\]
where $Ta = \dual{T}a = a$ and $Tb = \dual{T}b = b$.  The remainder of this section is focused on examining relationships among these 4-tuples.

Chain projection maps the interval $[a,b]_\T$ to two closed intervals on each of the four other chains, such as $[Pa, Pb]_\P$ and $[\dual{P}a, \dual{P}b]_\P$ along the chain $\P$.  Since we have shown that closed intervals on chains possess a unique scalar measure called \emph{length} (\ref{eq:length}) defined on the chain itself, this enables us to quantify the interval $[a,b]_\T$ with respect to the chain $\P$ by the pair of lengths $v_\P(Pb)-v_\P(Pa)$ and $v_\P(\dual{P}b)-v_\P(\dual{P}a)$ corresponding to the two closed intervals $[Pa, Pb]_\P$ and $[\dual{P}a, \dual{P}b]_\P$ on $\P$ onto which $[a,b]_\T$ projects. This 4-tuple $(v_\P(Pa), v_\P(Pb), v_\P(\dual{P}a), v_\P(\dual{P}b))_\P$ can be quantified by a pair of scalars $\big(v_\P(Pb)-v_\P(Pa), v_\P(\dual{P}b)-v_\P(\dual{P}a)\big)_\P$,
which we call the \emph{interval pair}.

Since the chains $\T$ and $\P$ are coordinated, we have that the closed interval $[a,b]_\T$ projects and back projects to closed intervals of equal length on $\P$, so that we can define $\Delta p = v_\P(Pb)-v_\P(Pa) = v_\P(\dual{P}b)-v_\P(\dual{P}a)$.  This results in a quantification with respect to the chain $\P$ consisting of an interval pair with equal components $(\Delta p, \Delta p)_\P$, which we refer to as a \emph{symmetric pair}.  Furthermore, the fact that each pair of chains is coordinated, implies that $\Delta p = \Delta r = \Delta t = \Delta s = \Delta q$, so that each of the chains in the coordinated set quantifies the closed interval with equal interval pairs.

The situation is not quite as straightforward for intervals in general.  We begin by considering quantification of a generalized interval with endpoints that are situated within the 1+1 dimensional subspace defined by a set of coordinated quantifying chains.  Later, Section \ref{sec:orthogonal_subspaces}, which is focused on orthogonal subspaces, is motivated by inconsistencies in quantification that arise when one or more endpoints of the interval being quantified are situated outside of this 1+1 dimensional subspace.  This results in a more general means of consistent quantification that is discussed in Section \ref{sec:dot-prod}.

Consider the quantification of a generalized interval by a single chain such that they are both situated within the same 1+1 dimensional subspace.  There are two possible cases to consider: (Case 1) the endpoints of the interval are both situated on the same side of the quantifying chain, and (Case 2) the endpoints of the interval are situated on opposite sides of the quantifying chain.  The specific example of the quantification of a closed interval situated on one of the chains of the coordinated set fully constrains the solution in Case 1, so that the interval is quantified by the chain $\P$ with the pair
$$\big(v_\P(Pb)-v_\P(Pa), v_\P(\dual{P}b)-v_\P(\dual{P}a)\big)_\P.$$

\begin{figure*}[t]
  \begin{center}
  \includegraphics[height=0.27\textheight]{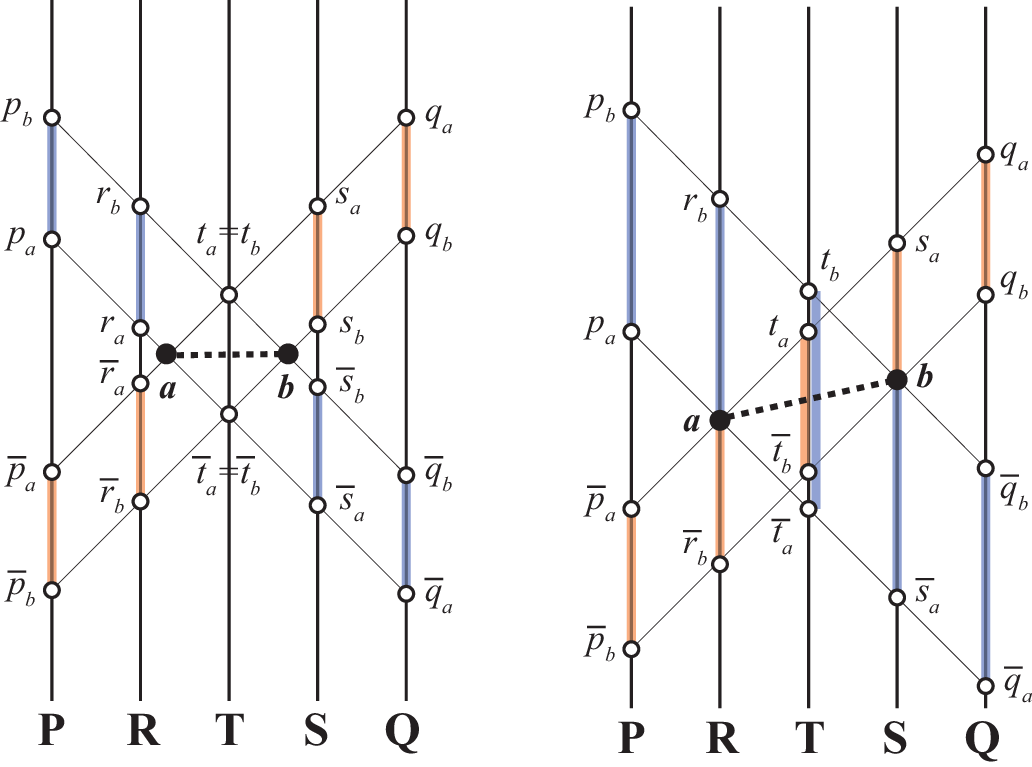}
  \end{center}
  \caption{(a) An illustration of the consistent quantification of a generalized interval $[a,b]$ by chains in a coordinated set of chains where the components of the interval pair are of opposite sign and equal in magnitude.  With respect to the coordinated set of chains, this interval is classified as a pure antichain-like interval.  (b) In general, antichain-like intervals are characterized by interval pairs where the components are of opposite sign.}
  \label{fig:pair-quantification-antichainlike}
\end{figure*}

We require that quantification in Case 2 be consistent with the quantification obtained in Case 1.  Consider the situation illustrated in \red Figures \ref{fig:pair-quantification-chainlike}b \black and \ref{fig:pair-quantification-antichainlike}a and b, where $\P | \T | \Q$ and $\P | a | \T | \Q$ and $\P | \T | b | \Q$ so that the endpoints of the interval $[a,b]$ are situated on the same side of chains $\P$ and $\Q$, but on opposite sides of $\T$, which we denote by $a | \T | b$.  Consistency of quantification by chains of the coordinated set requires that quantification by the chain $\T$ be equivalent to the quantification $\big(v_\P(Pb)-v_\P(Pa), v_\P(\dual{P}b)-v_\P(\dual{P}a)\big)_\P$ by the chain $\P$.
First consider the forward projections onto the chain $\P$.  The situation  $\P | a | \T$, which corresponds to Case II of (\ref{eq:collinearity_cases}), implies that
$$Pa = P\dual{T}a.$$
Similarly, the element $b$ satisfies Case III of (\ref{eq:collinearity_cases}), which results in
$$Pb = PTb.$$
We then have that
$$v_\P(Pb)-v_\P(Pa) = v_\P(PTb)-v_\P(P\dual{T}a),$$
and by the coordination condition,
$$v_\P(PTb)-v_\P(P\dual{T}a) = v_\T(Tb)-v_\T(\dual{T}a)$$
so that
$$v_\P(Pb)-v_\P(Pa) = v_\T(Tb)-v_\T(\dual{T}a).$$
Applying the same argument to the backward projections, we have that
$$v_\P(\dual{P}b)-v_\P(\dual{P}a) = v_\T(\dual{T}b)-v_\T(Ta),$$
so that in Case 2, one can quantify the interval with the pair $\big(v_\T(Tb)-v_\T(\dual{T}a), v_\T(\dual{T}b)-v_\T(Ta)\big)$.

The result is any generalized interval $[a,b]$ situated within a 1+1 dimensional subspace defined by a set of coordinated chains can be consistently quantified by any chain $\P$ in the coordinated set with the interval pair defined by
\begin{equation}
[a,b]\Big|_\P = \big(v_\P(Pb)-v_\P(\dual{P}a), v_\P(\dual{P}b)-v_\P(Pa)\big)_\P
\end{equation}
when $a | \P | b$ or $b | \P | a$, and by
\begin{equation}
[a,b]\Big|_\P = \big(v_\P(Pb)-v_\P(Pa), v_\P(\dual{P}b)-v_\P(\dual{P}a)\big)_\P
\end{equation}
when $\P | [a,b]$.
We can define
$$
\Delta p = v_\P(Pb)-v_\P(Pa)
$$
$$
\Delta \dual{p} = v_\P(\dual{P}b)-v_\P(\dual{P}a)
$$
so that we can conveniently write the interval pair in the case where the interval is on one side of the quantifying chain as
\begin{equation}
[a,b]\Big|_\P = \big(\Delta p, \Delta \dual{p}\big)_\P.
\end{equation}
In the case where the interval is situated between two chains, such as when $\P | a | \Q$ and $\P | b | \Q$, we can use the fact that chains $\P$ and $\Q$ are coordinated to write
$$v_\Q(Qb)-v_\Q(Qa) = v_\P(\dual{P}b)-v_\P(\dual{P}a)$$
so that the interval pair is expressed only in terms of forward projections.  The interval pair obtained by quantifying with respect to two chains, denoted $[a,b]\Big|_\PQ$, is given by
\begin{equation}
[a,b]\Big|_\PQ = \big(v_\P(Pb)-v_\P(Pa), v_\Q(Qb)-v_\Q(Qa))\big)_\PQ
\end{equation}
which is more conveniently written
\begin{equation}
[a,b]\Big|_\PQ = \big(\Delta p, \Delta q\big)_\PQ
\end{equation}
where $\Delta q = v_\Q(Qb)-v_\Q(Qa)$.
This is useful since the forward projections from $a$ and $b$ reflect information that observers, represented by chains $\P$ and $\Q$, may be able to access in a practical situation.

Last, this allows one to represent the length of a closed interval along a chain in terms of forward projections onto two coordinated observers.  Consider a closed interval $[p_j, p_k]_{\P}$ on the chain $\P$ of length $\Delta p$.  Since the chains $\P$ and $\Q$ are coordinated, one can write the length of the closed interval as
\begin{equation} \label{eq:length-coord}
d([p_j, p_k]_\P) \equiv \Delta p = \frac{\Delta p + \Delta q}{2},
\end{equation}
\red which we will find to be important later on. \black

\subsection{Quantifying Coordinated Chains}
\begin{figure}[t]
  \begin{center}
  \includegraphics[height=0.25\textheight]{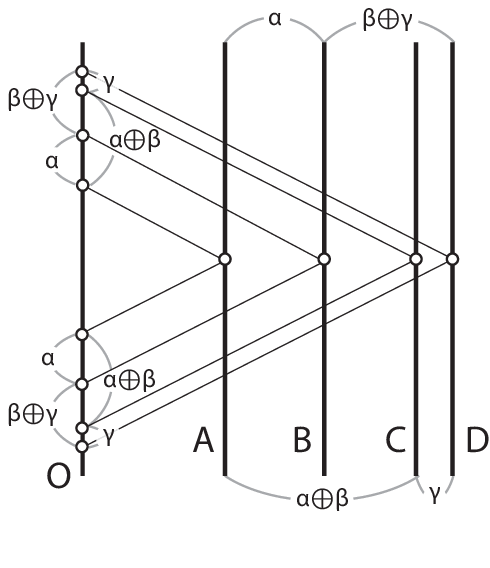}
  \end{center}
  \caption{We can quantify chains in a coordinated set of chains, just as we did events along a chain.  We can define an interval of chains as a set of chains between and including the two chains acting as endpoints, and define the distance to be a scalar quantity assigned to the interval.  For example $\INT{\A}{\C} = \{\A, \B, \C\}$.  The act of joining intervals of chains is associative, which leads to \red associativity of the distance measure, and thus additivity of the combinations of distances in (\ref{eq:additivity-of-distance}).}
  \label{fig:associativity-of-distance}
\end{figure}

In Section \ref{sec:induced-subspaces} we demonstrated that chains belonging to a set of coordinated chains can be ordered according to an ordering relation induced by chain projection, and that this gives rise to the concept of a 1+1 dimensional subspace.  We can therefore think of an abstract chain of coordinated chains, such as $\OO < \A < \B < \C < \D$ illustrated in Figure \ref{fig:associativity-of-distance}, and consider quantifying the coordinated chains themselves as well as intervals of chains---just as we have previously done for elements along a chain.

An \emph{interval of coordinated chains} can be defined as a set of coordinated chains between and including a pair of chains (endpoints) that belong to the coordinated set of chains.  For example, given the coordinated set of chains $\OO < \A < \B < \C < \D$, the interval of coordinated chains denoted by $\INT{\A}{\C}$ is given by the set of chains $\{\A, \B, \C\}$.  Two intervals of coordinated chains that share a single common chain can be joined via set union
\begin{equation}
\INT{\A}{\C} = \INT{\A}{\B} \setunion \INT{\B}{\C},
\end{equation}
which is analogous to joining closed intervals of events along a chain.
\red
We can assign scalar valuations to chains in the coordinated set, as well as to intervals of coordinated chains, which are themselves sets.  The result obtained in
Section \ref{sec:closed-intervals} applies, which requires quantification of intervals of chains to be additive (up to an invertible transform). So without loss of generality, we have that
\begin{equation} \label{eq:additivity-of-distance}
D(\INT{\A}{\C}) = D(\INT{\A}{\B}) + D(\INT{\B}{\C}).
\end{equation}
\black
We refer to the valuation $D(\INT{\A}{\B})$ as the \emph{distance} between chains $\A$ and $\B$.
Similarly, as described in Section \ref{sec:closed-intervals} for closed intervals of elements, additivity of $D$ results in
\begin{equation}
D(\INT{\A}{\C}) = X(\C) - X(\A),
\end{equation}
where $X(\A)$ and $X(\C)$ are the scalar valuations, referred to as \emph{position}, assigned to the chains $\A$ and $\C$.

\red Since quantification of the poset depends on the valuations assigned to the poset elements via chain projection, we require the distance between two chains in a coordinated set of chains to be a function of the valuations assigned to the elements on the two chains themselves and their projections onto one another.  That is, we seek a distance assignment that is a function of the interval pair associated with the generalized interval defined by two elements, one on each of the two chains.  Furthermore, since there can be no preferred elements, we require that the distance measure should not depend on which specific elements were chosen.  Thus given a pair of coordinated chains $\P$ and $\Q$ and arbitrarily chosen elements $p \in \P$ and $q \in \Q$ within the coordinated range, we consider the interval $[p,q]$ and require the distance between the chains $D(\INT{\P}{\Q})$ to be a function of the interval pair $(\Delta p, \Delta q)$ where $\Delta p = p - Pq$ and $\Delta q = Qp - q$.
\black

\begin{figure}[t]
  \begin{center}
  \includegraphics[height=0.20\textheight]{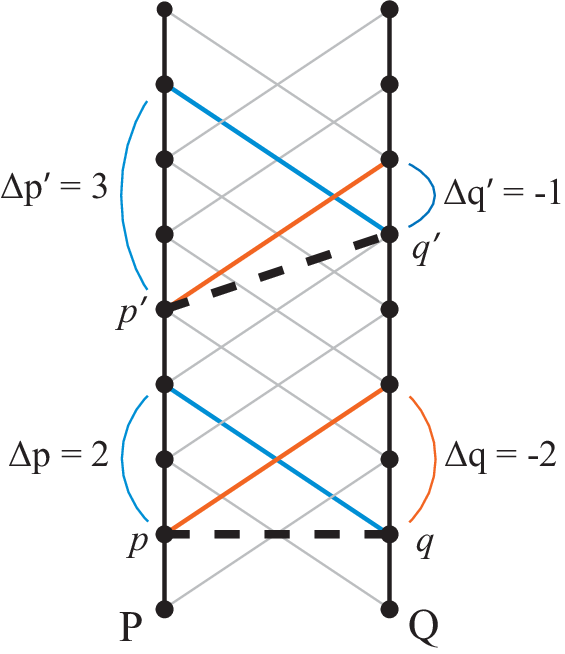}
  \end{center}
  \caption{The measure $D$ (distance) quantifying the relationship between two coordinated chains is a function of an element on each chain, but cannot depend on which elements were chosen. Intervals $[p,q]$ and $[p',q']$ are shown with distances $D([[P,Q]])$ given by $(\Delta p - \Delta q) / 2 = (2-(-2)) / 2 = 2$ and $(\Delta p' - \Delta q') / 2 = (3-(-1)) / 2 = 2$, respectively.}
  \label{fig:distance-measure}
\end{figure}

Associativity of joining intervals of coordinated chains is related to associativity of the join of projected closed intervals along a coordinated chain, as illustrated in Figure \ref{fig:associativity-of-distance}.  Since this results in additivity of distances as well as lengths of the projected intervals, this implies that the distance $D(\INT{\P}{\Q})$ must be a linear function of the elements of the interval pair $(\Delta p, \Delta q)_{\P\Q}$
\begin{equation}
D(\INT{\P}{\Q}) = a \Delta p + b \Delta q
\end{equation}
for arbitrary $p \in \P$ and $q \in \Q$.
\red Since distance is additive, the \black distance between a chain and itself \red must be \black zero, which implies that $a + b = 0$, so that $a = -b$.
\red Furthermore, the elements \black used to compute a distance should be \red arbitrary.  We \black confirm that one obtains the same distance using different elements \red by considering \black $p' \in \P$ and $q' \in \Q$ so that
\begin{equation}
a \Delta p + b \Delta q = a \Delta p' + b \Delta q'
\end{equation}
which can be rewritten as
\begin{equation}
a(p-Pq) + b(Qp-q) = a(p'-Pq') + b(Qp'-q')
\end{equation}
and rearranged to give
\begin{equation}
a(p-p') + b(q'-q) = a(Pq-Pq') + b(Qp'-Qp).
\end{equation}

Since the chains $\P$ and $\Q$ are coordinated, we have that
\begin{equation}
p'-p = Qp'-Qp,
\end{equation}
and
\begin{equation}
q'-q = Pq'-Pq.
\end{equation}
Substituting these into the expression above, we have that
\begin{equation}
a(p-p') + b(q'-q) = a(q-q') + b(p'-p).
\end{equation}
Since $p-p'$ and $q-q'$ are arbitrary, we have the condition that $a = -b$ so that
\begin{equation}
D(\INT{\P}{\Q}) = C(\Delta p - \Delta q),
\end{equation}
where $C$ is an arbitrary scale.  Without loss of generality, we can set $C = \frac{1}{2}$, which amounts to choosing units for distance that are consistent with the units chosen for length in (\ref{eq:length-coord}).  The resulting distance measure is
\begin{equation} \label{eq:distance}
D(\INT{\P}{\Q}) = \frac{\Delta p - \Delta q}{2}.
\end{equation}
As illustrated in Figure (\ref{fig:distance-measure}), this valuation is guaranteed to give the same distance from chain $\P$ to chain $\Q$ given any $p \in \P$ and $q \in \Q$ within the coordinated range.

\subsection{The Symmetric-Antisymmetric Decomposition} \label{sec:symmetric-antisymmetric}
These relationships suggest a useful decomposition, which we call the \emph{sym\-metric-anti\-sym\-metric decomposition}, where an interval pair $(\Delta p, \Delta q)$ is decomposed into the \red component-wise \black sum of a symmetric pair and an antisymmetric pair
\begin{widetext}
\begin{equation} \label{eq:symmetric-antisymmetric-pairs}
(\Delta p, \Delta q)_{\P\Q} =
\Big(\frac{\Delta p + \Delta q}{2},\frac{\Delta p + \Delta q}{2}\Big)_{\P\Q} + \\ \Big(\frac{\Delta p - \Delta q}{2},\frac{\Delta q - \Delta p}{2}\Big)_{\P\Q}.
\end{equation}
\end{widetext}
In this decomposition, the symmetric component represents projected lengths (\ref{eq:length-coord}) along the two coordinated chains, and the antisymmetric component represents a component of distance (\ref{eq:distance}) between the elements with respect to the subspace induced by the two coordinated chains.  We will examine this decomposition in more detail in Section \ref{sec:joining-generalized-intervals} when we look at joining generalized intervals.

\subsection{Interval Classes} \label{sec:IntervalClasses}
Given a 1+1 dimensional subspace induced by two coordinated chains $\P$ and $\Q$, the interval pair $\big(\Delta p, \Delta q\big)_\PQ$ enables us to identify three equivalence classes of generalized intervals $[a,b]$ situated within the 1+1 dimensional subspace induced by $\P$ and $\Q$ based on whether the two components of the pair are of like sign, opposite sign, or whether one of them is zero.

\begin{figure*}[t]
  \begin{center}
  \includegraphics[height=0.37\textheight]{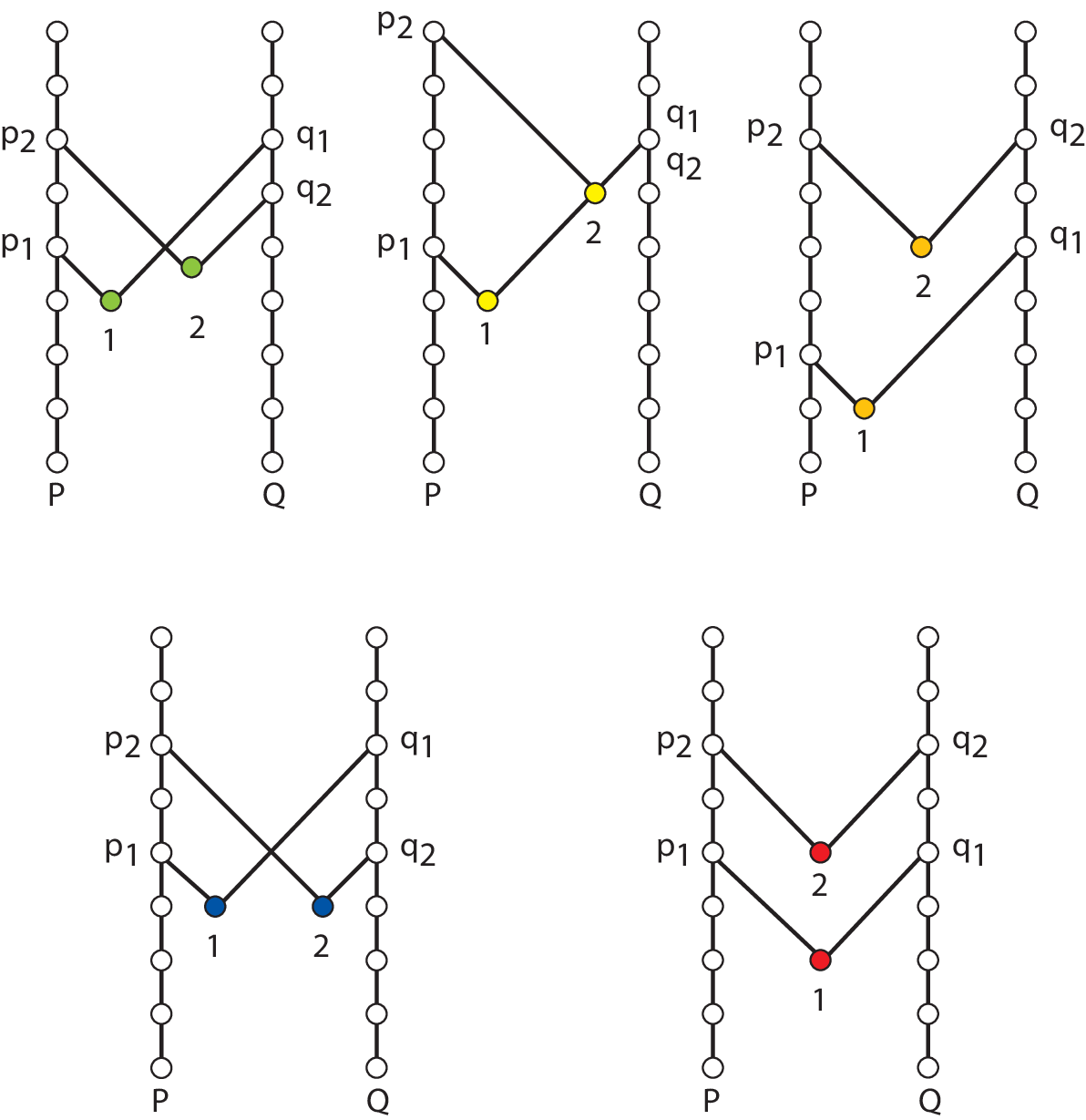}
  \end{center}
  \caption{This figure illustrates three classes of relationships between two elements forming a generalized interval and the observer chains.  (Top Left) \emph{Antichain-like} intervals have elements that project in opposite order to the two chains resulting in a pair quantification with opposite signs.  (Top Center) \emph{Projection-like} intervals are characterized by the fact that the two elements project to the same element on one of the observer chains so that one element of the quantifying pair is zero.  (Top Right) \emph{Chain-like} intervals are defined by elements that project to the observer chains in the same order resulting in a quantifying pair with like signs.  Note that since the elements may be incomparable, chain-like intervals do not necessarily lie on a chain.  However, all closed intervals along chains are chain-like.  (Bottom Left) A purely antichain-like interval is characterized by a quantifying pair that is anti-symmetric as in $(\Delta,-\Delta)$. (Bottom Right) A purely chain-like interval is quantified by a symmetric pair as in $(\Delta,\Delta)$.}
  \label{fig:event-relationships}
\end{figure*}

Closed intervals along a chain coordinated with $\P$ and $\Q$ are quantified by a \emph{symmetric pair}, such as $(\Delta p, \Delta p)$, where the components of the pair are both of like sign and equal magnitude.  Since these belong to the first class where the pair components are of like sign, we say that intervals in this equivalence class are \emph{chain-like} (see Figures \ref{fig:pair-quantification-chainlike} and \ref{fig:event-relationships}).  While all closed intervals along chains are chain-like, there exist generalized intervals with incomparable elements that do not form chains that are chain-like in the sense that the components of the interval pair are of the same sign.  Chain-like intervals quantified by a chain where each component of the interval pair is of equal magnitude (as in the case of a closed interval quantified by the host chain) are referred to as \emph{pure chain-like} as in Figure \ref{fig:pair-quantification-chainlike}a.  However, it should be noted that this latter classification is dependent on the relationship between the interval and the quantifying chain as it is not true in general that an interval characterized as pure chain-like by one chain will be characterized as pure chain-like by another chain.

Similarly, generalized intervals which are quantified by interval pairs with components of opposite sign are called \emph{antichain-like} (see Figures \ref{fig:pair-quantification-antichainlike} and \ref{fig:event-relationships}).  All antichain-like intervals are antichains \red where the endpoint elements are incomparable (not causally related).\black  In the case where the components of the interval pair are equal in magnitude, but opposite in sign, the interval is referred to as \emph{pure antichain-like} and the interval pair is said to be an \emph{antisymmetric pair} (see Figure \ref{fig:pair-quantification-antichainlike}a).  However, just as in the case of a pure chain-like interval, classification as a pure antichain-like interval is as dependent on the interval as it is the quantifying chain.

Generalized intervals that project to the same element on one of the two chains results in either $\Delta p = 0$ or $\Delta q = 0$.  In such cases, one element of the generalized interval projects to the other and then to one of the quantifying chains.  For this reason, we call such intervals \emph{projection-like} (see Figure \ref{fig:event-relationships}).

For a set of properly collinear chains, one can show that the elements of the interval pair for any interval cannot change sign when quantified by another pair of chains in the set.  We demonstrate this by contradiction.  Consider two properly collinear, but not coordinated, chains $\X$ and $\Y$ and assume that there exists an interval $[a,b]$ that projects to the closed interval $[Xa, Xb]$ on the chain $\X$ and the closed interval $[YXb, YXa]$ on the chain $\Y$ where $Xb > Xa$ and $YXa \geq YXb$ so that the projection changes sign.  Since the projection of $a$ onto a chain $\X$ is given by the least element of the chain $\X$ that includes $a$, we have that $YXa \geq YXb > Xb > Xa > a$ so that $YXb$ (not $YXa$) is the projection of $a$ onto $\Y$.  The result is that the closed interval $[Xa,Xb]$ projects either to $[YXa, YXb]$ (same order) or to $[YXb, YXb]$ (degenerate).  In the case where the order is preserved, the elements of the interval pair have the same sign, and in the degenerate case one of the elements is zero.  The result is that chain-like intervals can never be antichain-like intervals and vice versa, although either might be observed to be projection-like.

\subsection{Orthogonal Subspaces} \label{sec:orthogonal_subspaces}
In the previous sections, we showed how a pair of coordinated chains induces a 1+1 dimensional subspace in the poset.  Elements and chains not included in a given subspace may form subspaces of their own.  We begin by considering a particular example that motivates the concept of orthogonal subspaces.

Consider a pair of coordinated chains $\P$ and $\Q$, which form a subspace $\subspace{\P\Q}$, and a second pair of coordinated chains $\R$ and $\S$ not in $\subspace{\P\Q}$, which form a distinct subspace $\subspace{\R\S}$.  Consider $p \in \P$ and $q \in \Q$ such that they form a pure antichain-like generalized interval $[p,q]$ with respect to $\subspace{\P\Q}$, which when quantified results in
\begin{align}
[p,q]\Big|_{\P\Q} &= (\Delta,-\Delta)_\PQ\\
&= (v_\P(p)-v_\P(Pq), v_\Q(q)-v_\Q(Qp))_\PQ. \nonumber
\end{align}
where the notation $[p,q]\Big|_{\P\Q}$ indicates that the generalized interval $[p,q]$ is being quantified with respect to the chains $\P$ and $\Q$, which is also indicated by the subscript on the resulting pair.

\begin{figure*}[t]
  \begin{center}
  \includegraphics[height=0.30\textheight]{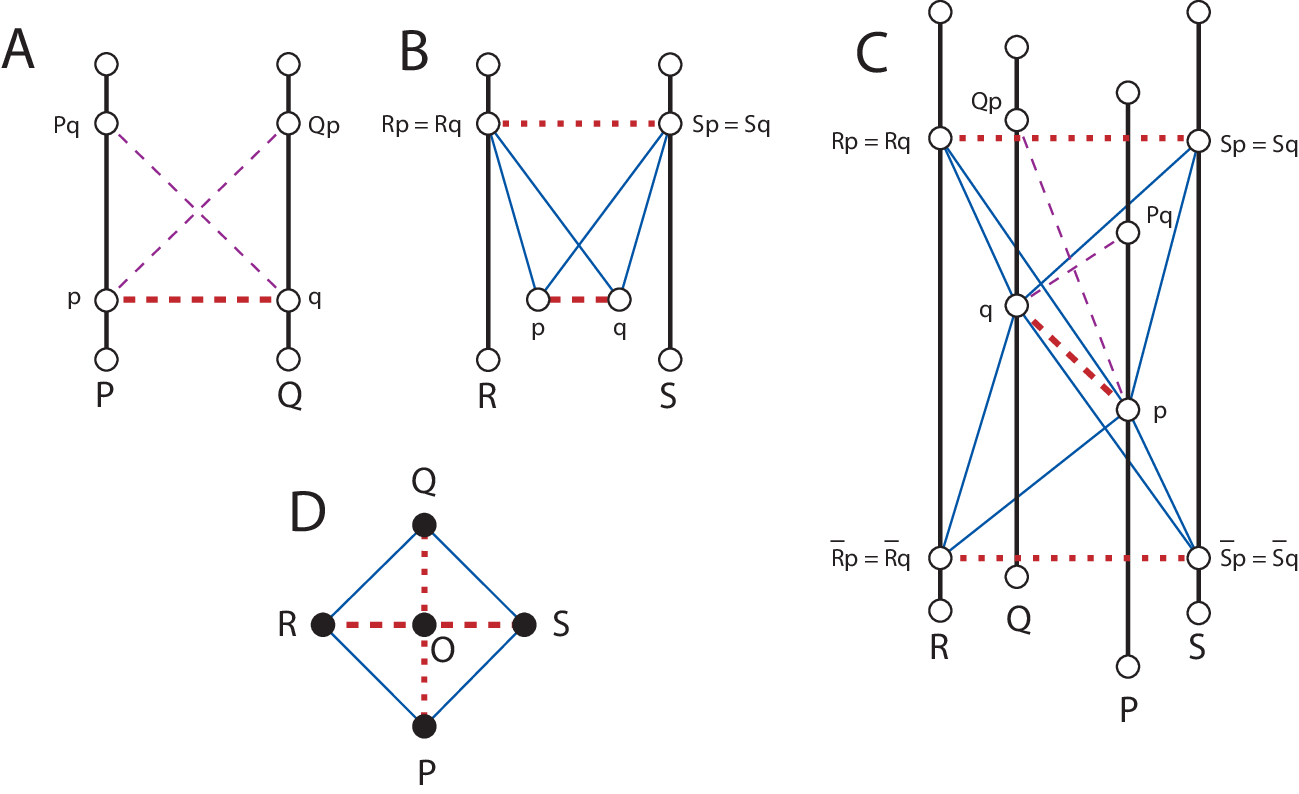}
  \end{center}
  \caption{(A) A pure anti-symmetric generalized interval $[p,q]$ in the subspace $\subspace{\P\Q}$. (B) Projection of $[p,q]$ onto $\R\S$ is quantified by the pair $(0,0)_{RS}$.  (C) An illustration of the relationship between the two subspaces where pure antichain-like intervals in $\subspace{\R\S}$ are quantified by $(0,0)_{PQ}$ \red with respect to \black $\subspace{\P\Q}$ and vice versa. (D) Illustrates the chains shown in (C) in a \emph{geometric view} where each chain is indicated by a dark circle and they are positioned relative to one another based on the betweenness relations.  The chain $\CHAIN{O}$ has been added so that it is properly collinear with both $\subspace{\P\Q}$ and $\subspace{\R\S}$ and situated between the two chains in each pair.}
  \label{fig:orthogonality}
\end{figure*}

Now consider that $p$ and $q$ project to $\R$ and $\S$ so that $Rp = Rq$ and $Sp = Sq$ resulting in a quantification of $[p,q]$ with respect to $\subspace{\R\S}$ equal to
\begin{align}
[p,q]\Big|_{\R\S} &= (v_R(Rp)-v_R(Rq),v_S(Sp)-v_S(Sq))_{\RS} \nonumber \\
&= (0, 0)_{\RS}.
\end{align}
This is illustrated in Figure \ref{fig:orthogonality}A-C.  Note also that the situation is similar for the antichain-like interval $[\overline{R}p, \overline{S}p]$ in $\subspace{\S\R}$ in that it is quantified by $(0,0)_\PQ$ with respect to $\subspace{\P\Q}$.  In this case we say that the subspaces $\subspace{\P\Q}$ and $\subspace{\S\R}$ are \emph{orthogonal} to one another.

Figure \ref{fig:orthogonality}D illustrates the chains shown in \ref{fig:orthogonality}(C) along with an additional chain $\CHAIN{O}$, which has been added so that it is properly collinear with both $\subspace{\P\Q}$ and $\subspace{\R\S}$.  We call this a \emph{geometric view} where each chain is indicated by a dark circle and they are positioned relative to one another based on the betweenness relations $\P|\CHAIN{O}|\Q$ and $\R|\CHAIN{O}|\S$.  The two independent 1+1 dimensional subspaces when combined results in a 2+1 dimensional subspace where the original ordering relation is supplemented with two ordering relations induced by the two independent properly collinear sets of chains.  The geometric view highlights the two induced dimensions while suppressing the original dimension along the chains.

While Figure \ref{fig:orthogonality}C illustrates a motivating example, the general situation is more subtle.  While an interval quantified by a pure antisymmetric pair with respect to the chains $\P$ and $\Q$ is quantified by $(0,0)$ with respect to the chains $\R$ and $\S$, that very same interval will not be quantified by $(0,0)$ with respect to the pair of chains $\R$ and $\CHAIN{O}$ or with respect to the pair of chains $\CHAIN{O}$ and $\S$.  In Section \ref{sec:dot-prod}, we derive a more advanced method of projection of an interval onto a subspace that yields consistent results when quantified with respect to any pair of chains in that subspace.

\subsection{Joining Generalized Intervals} \label{sec:joining-generalized-intervals}
In Section \ref{sec:closed-intervals}, we introduced the concept of joining closed intervals along a chain.  Here we extend this concept to that of joining generalized intervals.  Given two generalized intervals that share a \red single \black common element, such as $[a,b]$ and $[b,c]$, we can define a map $\concat$ that takes these two intervals to a third unique interval given by concatenation
\begin{equation} \label{eq:concatenation-of-intervals}
[a,c] = [a,b] \concat [b,c].
\end{equation}

In the case where each of these intervals is situated between a pair of coordinated chains $\P$ and $\Q$ in the $\subspace{\P\Q}$ subspace, the interval $[a,c]$ is quantified by the pair $(p_c-p_a,q_c-q_a)_{\P\Q}$, the interval $[a,b]$ by the pair $(p_b-p_a,q_b-q_a)_{\P\Q}$, and the interval $[b,c]$ by the pair $(p_c-p_b,q_c-q_b)_{\P\Q}$ so that the resulting interval pairs sum in a component-wise fashion in accordance with the \red concatenation operation (\ref{eq:concatenation-of-intervals}) \black
\begin{multline}
(p_c-p_a,q_c-q_a)_{\P\Q} = (p_c-p_b,q_c-q_b)_{\P\Q} \\
+ (p_b-p_a,q_b-q_a)_{\P\Q}.
\end{multline}

\begin{figure}[t]
  \begin{center}
  \includegraphics[height=0.20\textheight]{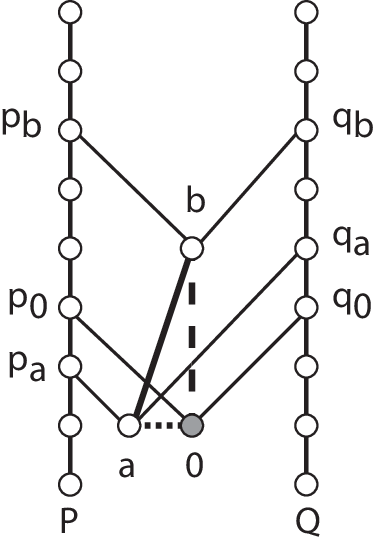}
  \end{center}
  \caption{This figure illustrates the symmetric-antisymmetric decomposition in the context of joining generalized intervals.  A generalized interval $[a,b]$ is quantified by the pair $(\Delta p, \Delta q) = (p_b-p_a, q_b-q_a)$.  This is equivalent to imagining an event $0$ defined by the projections $(p_0,q_0)$ onto $\P$ and $\Q$ such that $\Delta t = p_b-p_0 = q_b-q_0$ and \red $\Delta x = p_0-p_a = -(q_a-q_0)$ \black so that $[a,b] = [a,0] \concat [0,b]$, where $[a,0]$ is an antichain-like interval quantified by an antisymmetric pair $(\Delta x, -\Delta x)$ and $[0,b]$ is a chain-like interval quantified by a symmetric pair $(\Delta t, \Delta t)$.}
  \label{fig:symmetric-antisymmetric-decomposition}
\end{figure}

This enables us to decompose intervals by introducing an artificial event $0$ \red implicitly \black defined by a pair of elements $(p_0, q_0)$, which represent the forward projection of $0$ onto the pair of coordinated chains $\P$ and $\Q$.  Given two events $a$ and $b$ situated between two coordinated chains $\P$ and $\Q$, we can quantify the interval with the pair $(p_b-p_a, q_b-q_a)$.  Consider a decomposition
\begin{equation}
[a,b] = [a,0] \concat [0,b]
\end{equation}
where the pair $(p_0-p_a, q_0-q_a)$ quantifying $[a,0]$ is defined to be antisymmetric so that the two components of the pair are of equal magnitude but opposite sign, $p_0-p_a = q_a-q_0$, and the pair $(p_b-p_0, q_b-q_0)$ quantifying $[0,b]$ is defined to be symmetric so that the two components of the pair are of equal sign and magnitude $p_b-p_0 = q_b-q_0$.  These conditions are satisfied by
\begin{eqnarray}
p_0 &=& \frac{1}{2}(p_a + p_b + q_a - q_b) \\
q_0 &=& \frac{1}{2}(p_a - p_b + q_a + q_b).
\end{eqnarray}
The result is that any interval situated in the subspace defined by two coordinated chains can be expressed in terms of the join of a pure chain-like interval quantified by a symmetric pair and a pure antichain-like interval quantified by an antisymmetric pair (illustrated in Figure \ref{fig:symmetric-antisymmetric-decomposition}) in accordance with the symmetric-antisymmetric decomposition (\ref{eq:symmetric-antisymmetric-pairs})

One can also join intervals situated in distinct subspaces, although additivity of interval pairs does not hold.  For example, consider chains $\P$, $\CHAIN{O}$, and $\R$ in the situation illustrated in Figure \ref{fig:orthogonality}D and assume that they are pairwise coordinated.  Given elements $p \in \P$, $o \in \CHAIN{O}$, and $r \in \R$, we can construct intervals where
\begin{equation} \label{eq:join-orthog-intervals}
[p,r] = [p,o] \concat [o,r].
\end{equation}
Quantifying $[p,r]$ with respect to the coordinated pair of chains $\P$ and $\R$, we get the interval pair $(p_r-p, r-r_p)_{\P\R}$.  Quantifying the other two intervals within their respective subspaces, we obtain $(p_o-p,o-o_p)_{\P\CHAIN{O}}$ and $(o_r-o,r-r_o)_{\CHAIN{O}\R}$, where
\begin{multline}
(p_r-p, r-r_p)_{\P\R} \neq \\
(p_o-p,o-o_p)_{\P\CHAIN{O}} + (o_r-o,r-r_o)_{\CHAIN{O}\R}.
\end{multline}
Instead, we write
\begin{multline} \label{eq:pairs-under-concat}
(p_r-p, r-r_p)_{\P\R} \sim \\
(p_o-p,o-o_p)_{\P\CHAIN{O}} \concatplus (o_r-o,r-r_o)_{\CHAIN{O}\R}
\end{multline}
where $p=p_p$, $o=o_o$, $r=r_r$, and  the operator $\concatplus$, which is defined implicitly through the relation in (\ref{eq:join-orthog-intervals}), symbolically indicates a decomposition of the interval pair into distinct subspaces.

\subsection{Scalar Quantification of Intervals}  \label{sec:scalar-quantification-intervals}
We explore scalar measures of intervals, which will form the foundation for a metric.  As described above, closed intervals on a chain can be quantified by a scalar length.  To extend this concept of length to generalized intervals, we begin by identifying the constraints imposed on a scalar measure by the special case of a closed interval on one chain that forward projects and back projects onto two closed intervals on a second chain.

\begin{figure}[t]
  \begin{center}
  \includegraphics[height=0.30\textheight]{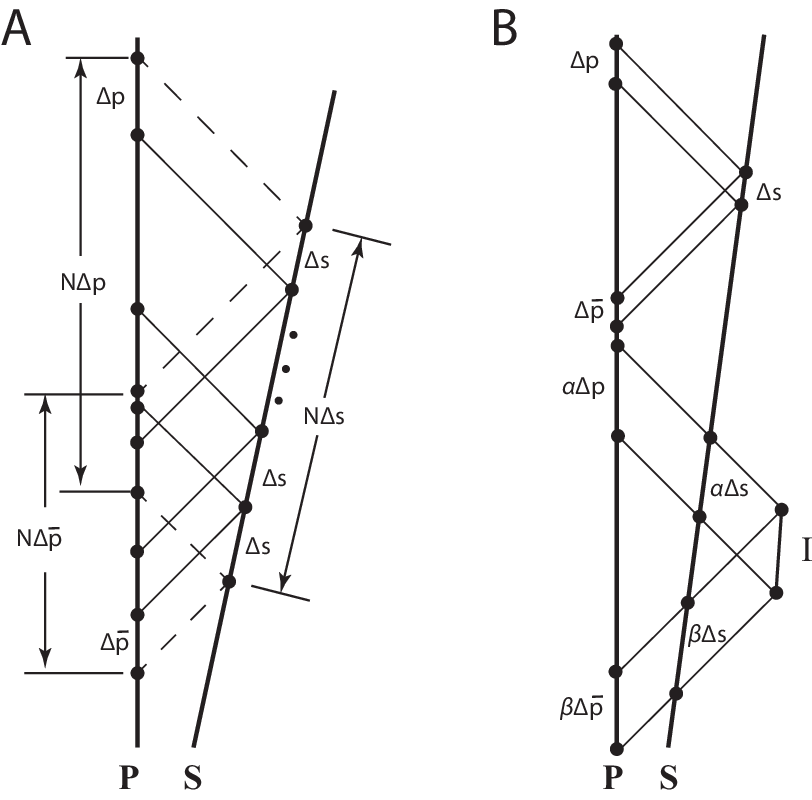}
  \end{center}
  \caption{(A) $N$ successive intervals each of length $\Delta s$ on the chain $\S$ forward project onto $N$ successive intervals each of length $\Delta p$ on $\P$ and back project onto $N$ successive intervals each of length $\Delta \overline{p}$.  (B) Projections of the chain-like interval $\I$ onto two linearly-related chains $\S$ and $\P$.}
  \label{fig:homogeneity}
\end{figure}

Consider a special case where two chains $\S$ and $\P$ that reside in the same 1+1D subspace are related such that $N$ successive intervals each of length $\Delta s$ on the chain $\S$ forward project onto $N$ successive intervals each of length $\Delta p$ on $\P$ and back project onto $N$ successive intervals each of length $\Delta \overline{p}$ as illustrated in Figure \ref{fig:homogeneity}A.  Note that in the special case where $\Delta p = \Delta \overline{p}$ we have that the chains are coordinated (up to scale) and that the lengths of the closed intervals are consistent with $\Delta p = \Delta \overline{p} = \Delta s$.  We refer to two such chains as being \emph{consistently-related} or, for reasons that will be made clear, \emph{linearly-related}.

Since each of these chains consistently quantifies closed intervals on the other, and the quantification of generalized intervals is performed by projecting onto these closed intervals, it is reasonable to require that the two chains consistently quantify all mutually quantifiable intervals.  That is, to obtain a unique scalar measure of an interval, we require that all linearly-related chains assign the same scalar (up to a common scale factor amounting to a choice of units) to all intervals that they both can quantify.  

Given a single closed interval on $\S$ with length $\Delta s$, and its quantification by $\P$ with the pair $(\Delta p, \Delta \overline{p})_\P$, we aim to identify a function $\sigma$, symmetric in its arguments, that takes the pair $(\Delta p, \Delta \overline{p})_\P$ to a scalar $\Delta s$ such that
\begin{equation}
\sigma(\Delta p, \Delta \dual{p}) = \Delta s.
\end{equation}
The scalar measures assigned to closed intervals on each of the two chains can be rescaled (change of units) and still remain consistent with additivity under the combination operator $\cup$ joining closed intervals along the chain.  Rescaling by a positive real number $\alpha$ results in
\begin{eqnarray}
\sigma(\alpha \Delta p, \alpha \Delta \dual{p}) &=& \alpha \Delta s, \\
&=& \alpha \sigma(\Delta p, \Delta \dual{p}).
\end{eqnarray}
This functional equation for $\sigma$ is of the general form known as the \emph{homogeneity equation}:
\begin{equation}
F(zx, zy) = z^k F(x,y)
\end{equation}
with $k=1$ whose solution can be written as \cite{Aczel:FunctEqns}
\begin{equation} \label{eq:homogeneity}
F(x,y) =
\begin{cases}
\sqrt{xy} ~ h(\frac{x}{y}) & \mbox{if } xy \neq 0  \\
ax & \mbox{if } x \neq 0, y = 0 \\
by & \mbox{if } y \neq 0, x = 0 \\
c & \mbox{if } y = 0, x = 0
\end{cases}
\end{equation}
where $h(\frac{x}{y}) = h(\frac{y}{x})$, since $\sigma$ and hence $F$, are symmetric in their arguments.
There is a symmetry between $x$ and $y$, so that $a = b$.  When $\Delta s = 0$ we have that $\Delta p = \Delta \overline{p} = 0$ so that $\sigma(0,0) = 0$ giving $c = 0$.
so that we can write
\begin{equation}
\sigma(x,y) = \label{eq:minimal-homogeneity}
\begin{cases}
\sqrt{xy} ~ h(\frac{x}{y}) & \mbox{if } xy \neq 0  \\
ax & \mbox{if } x \neq 0, y = 0 \\
ay & \mbox{if } y \neq 0, x = 0 \\
0 & \mbox{if } y = 0, x = 0.
\end{cases}
\end{equation}

By considering the special case where $\P$ and $\S$ are coordinated so that an interval of length $\Delta s$ on the chain $\S$ is quantified by $\P$ with the pair $(\Delta s, \Delta s)_\P$, we find that
\begin{equation}
\sigma(\Delta s ,\Delta s) = \Delta s ~ h(\frac{\Delta s}{\Delta s}) = \Delta s,
\end{equation}
which implies that $h(1) = 1$.

Having obtained a general form for the scalar measure, we consider the constraint imposed by a second special case.  Figure \ref{fig:homogeneity}B illustrates a chain-like interval $\I$ that projects onto two linearly-related finite chains $\S$ and $\P$ such that they are all properly collinear so that $\P|\S|\I$.  The closed interval of length $\Delta s$ on $\S$ is quantified by the pair $(\Delta p, \Delta \overline{p})_\P$ on $\P$.  If the interval $\I$ is quantified by the pair $(\alpha \Delta s, \beta \Delta s)_\S$ with respect to $\S$, then it is quantified by the pair $(\alpha \Delta p, \beta \Delta \overline{p})_\P$ with respect to $\P$.  Consistency requires that both chains assign the same scalar to the interval so that
\begin{equation}
\sqrt{(\alpha \Delta s)(\beta \Delta s)} ~ h\Big(\frac{\alpha \Delta s}{\beta \Delta s}\Big) =
\sqrt{(\alpha \Delta p)(\beta \Delta \overline{p})} ~ h\Big(\frac{\alpha \Delta p}{\beta \Delta \overline{p}}\Big),
\end{equation}
which can be simplified to
\begin{equation} \label{eq:step1}
\Delta s ~ h\Big(\frac{\alpha}{\beta}\Big) = \sqrt{\Delta p \Delta \overline{p}} ~ h\Big(\frac{\alpha \Delta p}{\beta \Delta \overline{p}}\Big).
\end{equation}
However, we have already established a relationship between chains $\P$ and $\S$ where closed intervals on $\S$ project to $\P$.  This amounts to setting $\alpha = \beta$ above so that
\begin{equation} \label{eq:k}
\Delta s = \sqrt{\Delta p \Delta \overline{p}} ~ h\Big(\frac{\Delta p}{\Delta \overline{p}}\Big).
\end{equation}
Substituting the expression for $\Delta s$ above into (\ref{eq:step1}), we find after some simplification that
$$
h\Big(\frac{\Delta p}{\Delta \overline{p}}\Big) h\Big(\frac{\alpha}{\beta}\Big) = h\Big(\frac{\alpha}{\beta} \frac{\Delta p}{\Delta \overline{p}}\Big)
$$
so that, letting $u = \frac{\Delta p}{\Delta \overline{p}}$ and $v = \frac{\alpha}{\beta}$ we have that
\begin{equation}
h(uv) = h(u)h(v).
\end{equation}

Since $h(1) = 1$, we can write
$$
h\Big(\frac{u}{u}\Big) = 1
$$
for non-zero $u$, which can be factored
$$
h(u) h\Big(\frac{1}{u}\Big) = 1.
$$
The function $h$ is symmetric in the sense that $h(\frac{u}{v}) = h(\frac{v}{u})$, which implies that $h(\frac{1}{u}) = h(u)$ and
$$
h(u)h(u) = 1,
$$
so that $h(u) = 1$ or $h(u) = -1$.  Equation (\ref{eq:k}) above gives an expression for $\Delta s$, which rules out $h(u) = -1$, so that $h(u) = 1$.

The result is that a chain-like interval quantified by the pair $(\Delta p, \Delta \dual{p})$ can be quantified by a unique scalar consistent among linearly-related chains given by
\begin{equation} \label{eq:sqrt-interval-scalar}
\sigma(\Delta p, \Delta \dual{p}) = \sqrt{\Delta p \Delta \dual{p}}.
\end{equation}
This special case constrains the functional form of this scalar, so that antichain-like intervals must be quantified similarly, but with $\Delta p \Delta \dual{p} \leq 0$, so that $\sqrt{\Delta p \Delta \dual{p}}$ is imaginary.  This suggests some kind of orthogonal relationship between pure chain-like intervals and pure antichain-like intervals, which are characterized by symmetric interval pairs and antisymmetric interval pairs, respectively.  This is further supported by the observation that, in the case of coordinated chains where $\Delta \bar{p} = \Delta q$, the argument of the square root, $\Delta p \Delta q$, can be written as
\begin{equation}
\Delta p \Delta q = \Big(\frac{\Delta p + \Delta q}{2}\Big)^2 - \Big(\frac{\Delta p - \Delta q}{2}\Big)^2,
\end{equation}
where the right-hand side of the equation above is a function of both the length along a chain (\ref{eq:length-coord}) and the distance between chains (\ref{eq:distance}). However, it remains to be shown that this is the \emph{only} reasonable scalar quantification.

To obtain a real-valued scalar quantification, which is some function $g(\Delta p, \Delta \dual{p})$, that applies to both chain-like and antichain-like intervals, we require that it be consistent with the result $\sigma(\Delta p, \Delta \dual{p}) = \sqrt{\Delta p \Delta \dual{p}}$ so that
\begin{equation}
g(\Delta p, \Delta \dual{p}) = F(\sigma(\Delta p, \Delta \dual{p})),
\end{equation}
where $F$ is an unknown function to be determined.
Consider the chain-like interval quantified by the pair $(\Delta t + \Delta x, \Delta t - \Delta x)$ where $\Delta t > \Delta x$.  We then have that
\begin{equation}
g(\Delta t + \Delta x, \Delta t - \Delta x) = F(\sqrt{(\Delta t + \Delta x)(\Delta t - \Delta x)}),
\end{equation}
which can be rewritten without loss of generality as
\begin{equation} \label{eq:big-G}
g(\Delta t + \Delta x, \Delta t - \Delta x) = G((\Delta t)^2-(\Delta x)^2)
\end{equation}
where $G$ is an unknown function to be determined.

We now rely on the fact that the joining of orthogonal intervals is \red associative, as must be the relationship among their interval pairs (\ref{eq:pairs-under-concat}) so that \black
\begin{multline}
((\Delta x, -\Delta x) \concatplus (\Delta y, -\Delta y)) \concatplus (\Delta z, -\Delta z) \\ = (\Delta x, -\Delta x) \concatplus ((\Delta y, -\Delta y) \concatplus (\Delta z, -\Delta z))
\end{multline}
where we have dropped the reference to the quantifying chains in each frame.  The corresponding scalar quantification is also associative
\begin{multline}
(g(\Delta x, -\Delta x) \,\hat{\concatplus}\, g(\Delta y, -\Delta y)) \,\hat{\concatplus}\, g(\Delta z, -\Delta z) \\ = g(\Delta x, -\Delta x) \,\hat{\concatplus}\, (g(\Delta y, -\Delta y) \,\hat{\concatplus}\, g(\Delta z, -\Delta z))
\end{multline}
where $\hat{\concatplus}$ is a function to be determined that combines the scalars of orthogonal intervals.  By defining
$u = g(\Delta x, -\Delta x)$, $v = g(\Delta y, -\Delta y)$, and $w = g(\Delta z, -\Delta z)$ we can rewrite the equation above as
\begin{equation}
(u \,\hat{\concatplus}\, v) \,\hat{\concatplus}\, w = u \,\hat{\concatplus}\, (v \,\hat{\concatplus}\, w).
\end{equation}
This functional equation for $\hat{\concatplus}$ is the \emph{associativity equation}, which is the very same functional equation (\ref{eq:associativity}) discussed in Section \ref{sec:closed-intervals} within the context of joining closed intervals along a chain.
The general solution $(\ref{eq:associativity-soln})$ is given by \cite{Aczel:FunctEqns}
\begin{equation}
u \,\hat{\concatplus}\, v = f(f^{-1}(u) + f^{-1}(v))
\end{equation}
where $f$ is an arbitrary invertible function.  This indicates that there exists a convenient representation where the scalar measures of orthogonal intervals are \red additive, and that we can adopt additivity without losing generality. \black

Pure chain-like intervals also enjoy associativity when joined with pure antichain-like intervals so the above theorem applies to them as well.  This implies that there exists an additive scalar obtained by applying $f^{-1} \circ g$ to the interval pair such that
\begin{multline} \label{eq:additivity-of-interval-scalar}
f^{-1}(g(\Delta t + \Delta x, \Delta t - \Delta x)) \\
= f^{-1}(g(\Delta t, \Delta t)) + f^{-1}(g(\Delta x, -\Delta x))
\end{multline}
and by applying (\ref{eq:big-G}) we have
\begin{multline}
f^{-1}(G((\Delta t)^2-(\Delta x)^2)) \\
= f^{-1}(g(\Delta t, \Delta t)) + f^{-1}(g(\Delta x, -\Delta x)).
\end{multline}
Since an interval quantified by the symmetric pair $(\Delta t, \Delta t)$ is chain-like, we have that
\begin{equation}
g(\Delta t, \Delta t) = G((\Delta t)^2),
\end{equation}
which allows us to write
\begin{multline}
f^{-1}(g(\Delta x, -\Delta x)) \\
= f^{-1}(G((\Delta t)^2-(\Delta x)^2)) - f^{-1}(G((\Delta t)^2)).
\end{multline}
Since the left-hand side can only depend on \red $\Delta x$, \black we have that $f^{-1} \circ G$ must be linear so that
\begin{equation}
f^{-1}(g(\Delta x, -\Delta x)) = -\lambda(\Delta x)^2
\end{equation}
and
\begin{equation}
f^{-1}(g(\Delta t, \Delta t)) = \lambda(\Delta t)^2
\end{equation}
where $\lambda$ is a constant which amounts to a freedom to select units. For additivity in (\ref{eq:additivity-of-interval-scalar}) to hold in general, we find that the constant $a=0$ in (\ref{eq:minimal-homogeneity}).

We can choose units consistent with those employed by the chain so that $\lambda=1$.  The function $f^{-1} \circ g$ represents a regraduated scalar quantification that is additive in the case of joining orthogonal intervals
\begin{equation}
f^{-1}(g(\Delta p, \Delta \dual{p})) = \Delta p \Delta \dual{p},
\end{equation}
which we rename with the composite symbol $\Delta s^2$
\begin{equation}
\Delta s^2(\Delta p, \Delta \dual{p}) = \Delta p \Delta \dual{p}
\end{equation}
to indicate its relationship to the scalar length $\Delta s$ obtained in the case of closed intervals along the quantifying chain. In the case where the interval being quantified is situated between two coordinated chains $\P$ and $\Q$ we have that $\Delta q = \Delta \dual{p}$ and can write
\begin{equation}
\Delta s^2(\Delta p, \Delta q) = \Delta p \Delta q.
\end{equation}
We call this scalar quantification of an interval, the \emph{interval scalar}.  It is additive when the intervals being joined are orthogonal, and its square root (or square root of its negation) is additive when the intervals being joined have the same ratio of symmetric and antisymmetric components.

Applying these results to the symmetric-antisymmetric decomposition (\ref{eq:symmetric-antisymmetric-pairs})
\begin{multline} \nonumber
(\Delta p, \Delta q)_{\P\Q} = \Big(\frac{\Delta p + \Delta q}{2},\frac{\Delta p + \Delta q}{2}\Big)_{\P\Q} \\
+ \Big(\frac{\Delta p - \Delta q}{2},\frac{\Delta q - \Delta p}{2}\Big)_{\P\Q},
\end{multline}
we verify that additivity of the interval scalar holds
\begin{equation} \label{eq:Minkowski-form}
\Delta s^2 = \Delta p \Delta q = \Big(\frac{\Delta p + \Delta q}{2}\Big)^2 - \Big(\frac{\Delta p - \Delta q}{2}\Big)^2.
\end{equation}
We refer to the resulting quadratic form on the right-hand side of (\ref{eq:Minkowski-form}) as the \emph{Minkowski form}, and stress that $\Delta s^2$ does not represent some quantity squared, but instead represents the product of two independent quantities $\Delta p$ and $\Delta q$.  \red This is the discrete version of the Minkowski metric. \black

\subsection{Pair Transformations}
\begin{figure}[t]
  \begin{center}
  \includegraphics[height=0.30\textheight]{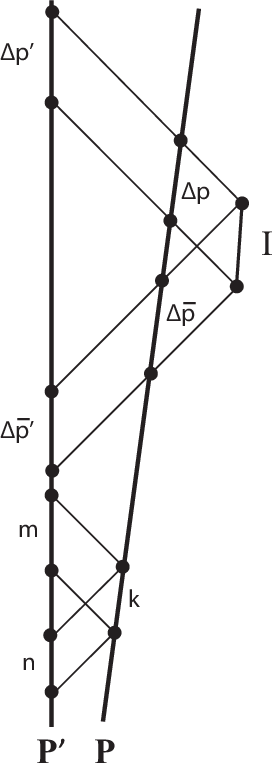}
  \end{center}
  \caption{This figure illustrates how the pair quantification of an interval $\I$ obtained using a chain $\P$ relates to the pair obtained by quantifying with respect to a linearly-related chain $\P'$.  The pair transformation can be written in terms of the transformation of a closed interval of length $k$ in $\P$ and its pair quantification by chain $\P'$.}
  \label{fig:pair-transform}
\end{figure}

Previously, we relied on linearly-related chains to constrain the form of the interval scalar.  Here we consider how the interval pair obtained by quantifying with respect to one chain transforms when quantified with respect to another linearly-related chain (Figure \ref{fig:pair-transform}).

A closed interval on $\P$ with length $k$ is quantified with respect to $\P$ by the symmetric pair $(k,k)_\P$.  This interval forward projects to a closed interval of length $m$ on $\P'$ and back projects to a closed interval of length $n$ on $\P'$ resulting in the pair $(m,n)_{\P'}$.  We write the pair transformation in terms of the function $L$ that takes one interval pair to another as
\begin{equation} \label{eq:basic-pair-transform}
L_{\P \rightarrow \P'}(k,k)_{\P} = (m,n)_{\P'}.
\end{equation}
The interval $\I$ projects onto $\P$ resulting in the pair $(\Delta p, \Delta \dual{p})_{\P}$, and also onto $\P'$ resulting in the pair $(\Delta p', \Delta \dual{p'})_{\P'}$ so that
\begin{equation}
L_{\P \rightarrow \P'}(\Delta p, \Delta \dual{p})_{\P} = (\Delta p', \Delta \dual{p'})_{\P'}.
\end{equation}
We can write the lengths of the projected intervals in terms of the length $k$ of the closed interval so that $\Delta p = \alpha k$, $\Delta \dual{p} = \beta k$, and $\Delta p' = \alpha m$, $\Delta \dual{p'} = \beta n$.  The pair transformation can then be written as
\begin{equation}
L_{\P \rightarrow \P'}(\alpha k, \beta k)_{\P} = (\alpha m, \beta n)_{\P'}
\end{equation}
indicating that the pair transformation is linear in each argument, as expected since the chains are linearly-related.  Constant terms in the linear transformation are zero since intervals of zero length always project to intervals of zero length.  Writing this linear transformation in terms of functions of $m$ and $n$, we find that in general
\begin{equation}
L_{\P \rightarrow \P'}(x, y)_{\P} = (x~f(m,n), y~g(m,n))_{\P'}
\end{equation}
where the functions $f$ and $g$ are to be determined.

The interval scalar associated with the closed interval allows us to write $k = \sqrt{mn}$ so that (\ref{eq:basic-pair-transform}) becomes
\begin{eqnarray}
L_{\P \rightarrow \P'}(k,k)_{\P} & = & (\sqrt{mn}~f(m,n), \sqrt{mn}~g(m,n))_{\P'} \nonumber \\
& = & (m,n)_{\P'}.
\end{eqnarray}
Equating components we find that
\begin{equation}
m = \sqrt{mn}~f(m,n)
\end{equation}
and
\begin{equation}
n = \sqrt{mn}~g(m,n)
\end{equation}
so that
\begin{equation}
f(m,n) = g^{-1}(m,n) = \sqrt{\frac{m}{n}}.
\end{equation}
Note that the fact that $f$ and $g$ are inversely related preserves the interval scalar in such situations.
In general, the pair transformation from one quantifying chain to another linearly-related chain is given by
\begin{eqnarray}
L_{\P \rightarrow \P'}(\Delta p, \Delta \dual{p})_{\P} &=& (\Delta p', \Delta \dual{p'})_{\P'}  \\
 &=& \Big(\Delta p \, \sqrt{\frac{m}{n}}, \Delta \dual{p} \, \sqrt{\frac{n}{m}}\Big)_{\P'}. \nonumber
\end{eqnarray}
where $m$ and $n$ are determined from (\ref{eq:basic-pair-transform}).

In the case of quantification by coordinated chains $\P$ and $\Q$, we can write the pair $(\Delta p, \Delta \dual{p})_{\P}$ as $(\Delta p, \Delta q)_{\P\Q}$ so that
\begin{eqnarray} \label{eq:pair-transform}
L_{\P \rightarrow \P'}(\Delta p, \Delta q)_{\P\Q} &=& (\Delta p', \Delta q')_{\P'\Q'} \\
 &=& \Big(\Delta p \, \sqrt{\frac{m}{n}}, \Delta q \, \sqrt{\frac{n}{m}}\Big)_{\P'\Q'}. \nonumber
\end{eqnarray}

The fundamental nature of the pair of projections is manifest in the simplicity of this transformation.  We observe that this is related to the Bondi k-calculus \cite{Bondi:1980} \red formulation of the Lorentz transformations, \black
as well as Kauffman's iterant algebra \cite{Kauffman:1985} which treats the transformation as a pair-wise multiplication.

\section{The Space-Time Picture} \label{sec:spacetime-picture}
In this section, we introduce a change of variables motivated by the identification of three distinct classes of intervals induced by chain projection.  We demonstrate that this change of variables results in a metric analogous to the Minkowski metric and reveals that the pair transformation is analogous to a Lorentz transformation in an analogous space-time.

\subsection{Space-Time Coordinates}
The symmetric-antisymmetric decomposition suggests a convenient change of variables:
\begin{eqnarray}
\Delta t &=& \frac{\Delta p + \Delta q}{2} \\
\Delta x &=& \frac{\Delta p - \Delta q}{2}
\end{eqnarray}
where
\begin{eqnarray}
\Delta p &=& \Delta t + \Delta x \\
\Delta q &=& \Delta t - \Delta x.
\end{eqnarray}
With this definition, any interval pair $(\Delta p, \Delta q)$ can be written as
\begin{equation}
\Delta s^2 = (\Delta p, \Delta q) = (\Delta t, \Delta t) + (\Delta x, -\Delta x),
\end{equation}
where we refer to the two pairs on the right as the time and space components, respectively.  Similarly, the interval scalar can be written as
\begin{equation}
\Delta p \Delta q = {\Delta t}^2 - {\Delta x}^2
\end{equation}
which is analogous to the Minkowski metric of flat space-time.

There are a few important observations to make at this point.  First, the interval scalar originates from the product of a pair of quantities and is not some fundamental quantity squared, as is suggested by the usual notation ${\Delta s}^2$.  Second, while this derivation suggests that the Minkowski metric may always be employed by a chain to quantify intervals, this does not mean that it is the most convenient description.  That is, if the time and space components of a sequence of intervals \red have the same value \black when quantified with respect to one chain, then, in general, it is not true that \red they will have the same value \black when quantified by another chain.  That is, one may always use this metric, but lengths and times \red observed to be constant by one chain may vary with respect to a second chain. \black

The time and space coordinates can be applied to the pair transformation (\ref{eq:pair-transform})
\begin{equation}
(\Delta p', \Delta q') = \Big(\Delta p \, \sqrt{\frac{m}{n}}, \Delta q \, \sqrt{\frac{n}{m}}\Big).
\end{equation}
Changing variables from $\Delta p$ and $\Delta q$ to coordinates $\Delta t$ and $\Delta x$, mixes the pair resulting in a linear transformation
\begin{multline}
\Big(\Delta t' + \Delta x', \Delta t' - \Delta x' \Big) = \\
\Big((\Delta t + \Delta x) \sqrt{\frac{m}{n}}, (\Delta t - \Delta x) \sqrt{\frac{n}{m}} \Big),
\end{multline}
which can be represented by a matrix multiplication. Solving for
$\Delta t'$ and $\Delta x'$, we find that
\begin{eqnarray}
\Delta t' & = & \frac{\sqrt{\frac{m}{n}} + \sqrt{\frac{n}{m}}}{2} \Delta t + \frac{\sqrt{\frac{m}{n}} - \sqrt{\frac{n}{m}}}{2} \Delta x \\
\Delta x' & = & \frac{\sqrt{\frac{m}{n}} - \sqrt{\frac{n}{m}}}{2} \Delta t + \frac{\sqrt{\frac{m}{n}} + \sqrt{\frac{n}{m}}}{2} \Delta x.
\end{eqnarray}

By defining
\begin{equation} \label{eq:beta}
\beta = \frac{m-n}{m+n},
\end{equation}
we obtain a relation analogous to the \emph{Lorentz transformation} in coordinate form
\begin{eqnarray}
\Delta t' & = & \frac{1}{\sqrt{1-{\beta}^2}} {\Delta t} + \frac{-{\beta}}{\sqrt{1-{\beta}^2}} {\Delta x} \\
\Delta x' & = & \frac{-{\beta}}{\sqrt{1-{\beta}^2}} \Delta t + \frac{1}{\sqrt{1-{\beta}^2}}  \Delta x,
\end{eqnarray}
which can be further simplified by defining $\gamma = \frac{1}{\sqrt{1-{\beta}^2}}$ and writing the linear transformation as a matrix multiplication
\begin{equation}
 \left[ \begin{array}{c} \Delta t' \\ \Delta x' \end{array} \right] =
 \begin{bmatrix} \gamma & -{\beta}\gamma \\ -{\beta}\gamma & \gamma \end{bmatrix}
 \left[ \begin{array}{c} \Delta t \\ \Delta x \end{array} \right].
\end{equation}
These results suggest that time and space can be viewed in terms of a uniquely consistent means of quantifying intervals.

\subsection{Motion}
The quantity
\begin{equation}
\beta = \frac{m-n}{m+n},
\end{equation}
introduced in the derivation above
is the relevant quantity that relates two linearly-related chains that project to one another in a constant fashion.  Its dependence on the antisymmetric component of the pair results in its antisymmetric behavior when the chains are interchanged.  That is, $\beta$ is antisymmetric in the sense that if chain $\P$ is related to chain $\R$ by $\beta$, then chain $\R$ relates to chain $\P$ by $-\beta$.  In the special case where $m=n$, we have that $\beta = 0$.  This situation represents two coordinated chains, which in the space-time picture is analogous to two observers at rest with respect to one another.

Moreover, this quantity has the extreme values of $\beta = \pm 1$ which correspond to the cases where $m=0$ and $n=0$ indicating that all elements of one chain project onto the same element of the other resulting in an interval scalar of zero.  This is a degenerate situation in the sense that one chain can quantify the other, but not vice versa.  Such intervals are classified as projection-like, which in the space-time picture are analogous to light-like intervals.  Since the interval scalar is invariant among linearly-related chains, we have that if $\beta$ is extremal with respect to one chain, then it must be extremal with respect to all other linearly-related chains.  This is analogous to the experimentally observed fact that the speed of light, which is the maximum speed, is invariant for all inertial frames.

When comparing three or more chains that project to one another in a constant fashion, the values of $\beta$ describing the relationship between pairs of such chains are related by the familiar velocity addition rule, which can be derived from this point as a standard exercise.

It is important to keep in mind that $\beta$ represents a relationship between chains in a partially-ordered set.  There is no motion in a partially-ordered set---only connectivity.  This suggests that physical motion can be interpreted in terms of connectivity, or equivalently that motion is a manifestation of interaction.

\subsection{Coordinates and the Pythagorean Decomposition}

Given the form of the interval scalar, and the fact that it is additive for orthogonal intervals, we find that this leads immediately to the Pythagorean theorem.

Consider the two orthogonal subspaces defined by the coordinated chains $\P|\CHAIN{O}|\Q$ and $\R|\CHAIN{O}|\S$ as illustrated in Figure \ref{fig:orthogonality}. Consider three events $p \in \P$, $o \in \CHAIN{O}$ and $r \in \R$ such that they define three pure antisymmetric intervals $A = [p,o]$, $B = [o,r]$ , and $C = [p,r]$ so that
\begin{equation}
[p,r] = [p,o] \uplus [o,r]
\end{equation}
and
\begin{multline}
(p_r-p, r-r_p)_{\P\R} \sim \\
(p_o-p,o-o_p)_{\P\CHAIN{O}} \oplus (o_r-o,r-r_o)_{\CHAIN{O}\R}
\end{multline}
where $\Delta a = p_o-p = -(o-o_p)$, $\Delta b = o_r-o = -(r-r_o)$, and $\Delta c = p_r-p = -(r-r_p)$ so that we can write
\begin{equation}
(\Delta c, -\Delta c)_{\P\R} \sim (\Delta a, -\Delta a)_{\P\CHAIN{O}} \oplus (\Delta b, -\Delta b)_{\CHAIN{O}\R}.
\end{equation}
Since the intervals $A$ and $B$ are orthogonal, the interval scalars sum resulting in
\begin{equation} \label{eq:pythagorean}
{\Delta c}^2 = {\Delta a}^2 + {\Delta b}^2,
\end{equation}
which is the familiar Pythagorean theorem applied to purely space-like intervals.

This enables one to quantify an interval with respect to an extant set of orthogonal subspaces based on projections of that interval onto the chains in those subspaces thus defining a multidimensional coordinate system.  This can be made more abstract by defining imaginary coordinate axes based on imagined, but consistent, projections.  This is accomplished by introducing an imaginary decomposition chain $\CHAIN{O}$ where the projections of an interval onto this chain are parameterized by a single parameter $\theta$.  As an example, consider an interval $[p_a,r_b]$ defined by $p_a \in \P$ and $r_b \in \R$.  This interval is quantified by the pair $(p_a-Pr_b, r_b-Rp_a)_{\P\R}$, which can be decomposed into symmetric and antisymmetric pairs
\begin{multline}
(p_a-Pr_b, r_b-Rp_a)_{\P\R} =\\
 (\Delta t, \Delta t)_{\P\R} + (\Delta r, -\Delta r)_{\P\R}.
\end{multline}
We can then introduce the chain $\CHAIN{O}$ where, introducing the convenient \emph{dot-star notation} where $(\Delta t, \cdot) \doteq (\Delta t, \Delta t)$ and $(\Delta r, *) \doteq (\Delta r, -\Delta r)$, we can write
\begin{multline}
(\Delta t, \cdot)_{\P\R} + (\Delta r, *)_{\P\R} \sim \\
(\Delta t, \cdot)_{\P\R} \oplus (\Delta r~f(\theta), *)_{\P\CHAIN{O}} \oplus (\Delta r~g(\theta), *)_{\CHAIN{O}\R}
\end{multline}
with the condition that $f^2(\theta) + g^2(\theta) = 1$ to preserve the interval scalar.
There are many such parameterizations one could
choose.  For the purpose of illustration, we will choose
$f(\theta) = \sin \theta$ and $g(\theta) = \cos \theta$ so that
\begin{multline}
(\Delta t, \cdot)_{\P\R} + (\Delta r, *)_{\P\R} \sim \\
(\Delta t, \cdot)_{\P\R} \oplus (\Delta r \sin \theta, *)_{\P\CHAIN{O}} \oplus (\Delta r \cos \theta, *)_{\CHAIN{O}\R},
\end{multline}
and
\begin{equation}
\Delta t^2 - \Delta r^2 \; = \; \Delta t^2 - \Delta r^2 \sin^2 \theta  - \Delta r^2 \cos^2 \theta.
\end{equation}
The antisymmetric pair $(\Delta r \sin \theta, *)$ can be further
decomposed by introducing an additional decomposition chain $Q'$ along
with an accompanying parameterization $\phi$, so that
\begin{multline}
(\Delta r \sin \theta, *)_{\P\CHAIN{O}} \sim \\
(\Delta r \sin \theta \, f(\phi), *)_{\P\Q} \oplus (\Delta r \sin \theta \, g(\phi), *)_{\Q\CHAIN{O}},
\end{multline}
where again $f^2(\phi) + g^2(\phi) = 1$. Choosing $\sin$ and
$\cos$ as before, and rearranging terms, results in the pair decomposition
\begin{multline}
(\Delta t, \cdot)_{\P\R} + (\Delta r, *)_{\P\R} \; \sim
(\Delta t, \cdot)_{\P\R} \\
\oplus \; (\Delta r \sin \theta \, \cos \phi, *)_{\P\Q} \\
\oplus \; (\Delta r \sin \theta \, \sin \phi, *)_{\Q\CHAIN{O}} \\
\oplus \; (\Delta r \cos \theta, *)_{\CHAIN{O}\R},
\end{multline}
and the interval scalar
\begin{multline}
\Delta t^2 - \Delta r^2 \;  =
\Delta t^2 \; - \; \Delta r^2 \sin^2 \theta  \, \cos^2 \phi \; \\
- \; \Delta r^2 \sin^2 \theta  \, \sin^2 \phi \; - \; \Delta r^2 \cos^2 \theta,
\end{multline}
which is the familiar representation of the Minkowski metric in
spherical coordinates $(r, \theta, \phi)$.  A
change of variables to $z = r \cos \theta$ and $\rho = r \sin
\theta$ results in a representation of the metric in cylindrical
coordinates, and a further change to $x = r \sin \theta \,  \cos
\phi$ and $y = r \sin \theta \,  \sin \phi$ results in a
representation of the metric in three-dimensional Cartesian
coordinates
\begin{equation}
\Delta t^2 - \Delta r^2 \; = \; \Delta t^2 - \Delta x^2  - \Delta y^2 - \Delta z^2.
\end{equation}

\section{Subspace Projection} \label{sec:dot-prod}
We have seen that intervals can only be consistently quantified via chain projection if they are situated within the subspace defined by the quantifying chains.  Here we introduce a more generally useful projection method that results in consistent quantification of intervals by chains in a subspace when using \emph{any pair} of chains in a coordinated set of chains defining a subspace.

\begin{figure*}[t]
  \begin{center}
  \includegraphics[width=0.40\textheight]{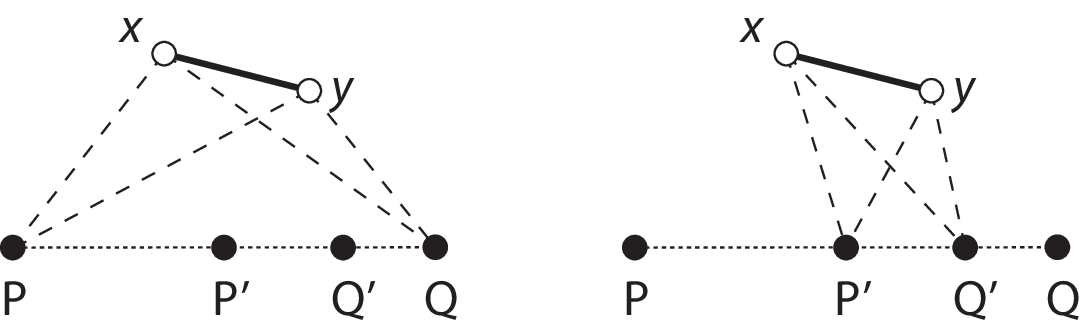}
  \end{center}
    \caption{This figure illustrates the subspace projection of an interval onto a pair of coordinated chains.  It is a more advanced projection method in the sense that it provides consistent quantification of the interval by any pair of chains in the coordinated set of chains defining the subspace. The contingencies, such as which chains in the subspace are used for quantification, are eliminated by the antisymmetric combination of terms.}
  \label{fig:subspace-projection}
\end{figure*}

Given an element $x$ and a chain $\P$, which is one of several in a coordinated set of chains, one can compute the distance between the element and the chain as an antisymmetric combination of projections
\begin{equation}
d(x,\P) = \frac{(p-Px)-(\dual{P}p-\dual{P}x)}{2},
\end{equation}
\red where $p$ is an arbitrary element in $\P$. \black
Similarly, we define the projection of the interval $[x,y]$ onto the subspace $\subspace{\P\Q}$ in terms of the antisymmetric combination of the squares of the four relevant distances
\begin{equation}
({d(y,\P)}^2 - {d(x,\P)}^2) - ({d(y,\Q)}^2 - {d(x,\Q)}^2)
\end{equation}
which can be rewritten as
\begin{equation}
({d(y,\P)}^2 - {d(y,\Q)}^2) - ({d(x,\P)}^2 - {d(x,\Q)}^2).
\end{equation}
However, this quantity depends on the distance between the chains $\P$ and $\Q$, $d(\P,\Q)$, defined in (\ref{eq:distance}).  Using the Pythagorean theorem (\ref{eq:pythagorean}) one can show that to obtain a consistent quantification of the interval by any pair of coordinated chains within the subspace, we simply normalize $\Pi_{\P\Q} [x,y]$ by twice the distance from $\P$ to $\Q$ resulting in
\begin{multline}
\Pi_{\subspace{\P\Q}} [x,y] = \\
\frac{({d(y,\P)}^2 - {d(y,\Q)}^2) - ({d(x,\P)}^2 - {d(x,\Q)}^2)}{2 d(\P,\Q)}.
\end{multline}

By combining the projections of the interval onto two chains in this way, we have developed a more advanced method of projection, which enables one to quantify any generalized interval consistently with respect to a given subspace.  This method, which we call \emph{subspace projection}, is the poset analogue of the \emph{inner product} or \emph{dot product}.  Moreover, it is important to note that it is the antisymmetric combination of squared distances that eliminates contingencies, such as which two chains in the subspace are used to perform the quantification.

\section{Conclusion}
We have considered a simple picture of interactions that focuses only on the fact that particles influence one another.  Events are defined as the boundaries of influence with one event representing the act of influencing and the other event representing the corresponding reaction.  The result is a partially-ordered set, or a poset, of events.

\red Consistent quantification \black of ordered structures, such as posets, by assigning n-tuples of numbers (real or natural) to elements or sets of elements, such as intervals, leads to faithful mathematical representations of that structure.  In cases where ordered sets (or related algebraic structures) possess symmetries, these symmetries will place constraints on any proposed quantification scheme resulting in numeric constraint equations, which can be identified as laws \cite{Knuth:laws}.  This has been demonstrated previously in the case of Boolean, and the more general distributive lattices \cite{Knuth:measuring}.  In the case of a Boolean lattice of logical statements, the sum and product rules of probability theory emerge as constraint equations \cite{Knuth&Skilling:2012}.  Similarly, \red consistent quantification \black using pairs of numbers applied to the composition of measurement sequences is constrained by the algebraic relations of combining measurements in series and in parallel.  In this case, the symmetries of the algebraic relations result in constraint equations that have been shown to be equivalent to Feynman's sum and product rules for  quantum mechanical amplitudes \cite{GKS:PRA}\cite{GK:Symmetry}.

However, the fact that posets lack general symmetries means that previously-developed symmetry-based methods of \red consistent quantification \black cannot be directly applied to posets in general.  Here we have shown that the identification of one or more distinguished chains in a poset induces sufficient symmetry to impose useful constraints on quantification and that these constraints in the case of coordinated chains lead directly to the mathematics of special relativity.  Specifically, the interval scalar, which represents the unique quantification of intervals (up to a scale) represents the poset analogue of the Minkowski metric.  In addition, quantification of an interval with respect to two linearly-related pairs of chains results in two interval pairs that are related by a pair transformation, which is shown to be equivalent to a Lorentz transformation.  This enables one to adopt a space-time perspective, which focuses on the chain-like and antichain-like symmetries induced by the distinguished chain/s at the expense of the simplicity of the mathematics inherent to the poset picture.

Derivation of the Lorentz transformations from causality and fundamental symmetries is not without precedent.  Although most past approaches either assume a Minkowski metric, or at the very least, the existence of a space-time manifold endowed with a metric.  Zeeman \cite{Zeeman:1964} showed that representing causality as a partial ordering on a Minkowski space forces the Lorentz group.  Levy-LeBlond \cite{Levy-LeBlond:1976} produced another derivation that results in both the Lorentz transformation and the Galilean transformation, which does not rely on the Minskowki metric, but rather on the homogeneity and isotropy of space and causality.  Kauffman \cite{Kauffman:1985} takes this further by considering the principle of relativity as invariance under linear transformations and derives the Lorentz transformation, Galilean transformation, and rotation as special cases.  We had been aware of Kauffman's results and the fact that such symmetries result in linear transformations, but after completing our work we were impressed by the similarities between his iterant coordinates and our interval pair, both of which transform in accordance with Bondi's k-calculus \cite{Bondi:1980}.  The common element in each of these approaches is the concept of radar time, which manifests itself in our poset approach via chain projection. There has also been a great deal of work to derive space-time geometry (e.g. \cite{Hawking+etal:1976}\cite{Malament:1977}).  Of particular note due to its similarity to the present effort is the work by Ehlers, Pirani and Schild \cite{Ehlers-etal:1972} who accomplish this by considering properties of light signals along with the assumption of the existence of a space-time manifold.  This enables them to extend their results to general relativity by deriving the properties of a curved space-time that is assumed to exist.

After submission of an early version of this
paper to the arXiv \cite{Knuth+Bahreyni:SR}, we were introduced to the work of
D'Ariano \cite{D'Ariano:2010} who showed how the Lorentz
transformations can be derived, in principle, from event-counting
performed by an observer within a causal network implemented by a
quantum computer. This has since been worked out in more detail,
and most of the quantum mechanical framework has been abstracted
away \cite{D'Ariano&Tosini:2010}.  D'Ariano's approach is similar
in spirit to ours in that causality plays a central role, however
it relies on a homogeneous causal network, which is unphysical, and focuses on \red deriving the \black
Lorentz transformation rather than the Minkowski metric.

In contrast to previous approaches, the approach presented here provides additional insights.  The antisymmetry of space arises from the antisymmetry of the projections of the antichain-like intervals.  As such, antichain-like intervals can be further decomposed into the join of orthogonal intervals via the Pythagorean theorem, whereas chain-like intervals enjoy no such decomposition.  Another way to look at this is to consider that the distinguished chain, which represents a total order, gives rise to one-dimensional time.  Sets of multiple coordinated chains induce a different kind of ordering relation, which gives rise to spatial dimensions, which can be multiple in number.  Unfortunately, the symmetries introduced here by a distinguished chain are insufficient to constrain the number of spatial dimensions.  This can be demonstrated by construction.  Consider $N$ finite chains labeled by $C_i$ where $1 \leq i \leq N$.  Let each chain $C_i$ consist of two events $x_i < y_i$, where the index $i$ indicates chain $C_i$ to which the element belongs.  Let $y_i \geq x_j$ for all $1 \leq i \leq N$ and $1 \leq j \leq N$ so that $y_i$ includes each of the $x$ events on each of the $N$ chains.  By symmetry, the distance between each pair of chains in the set of $N$ chains is equal, and thus is only possible in an $(N-1)$ dimensional space.  So the present approach admits one-dimensional time and multi-dimensional space.  Physics presumably enters by dictating the allowed connectivity of such a poset, and this is expected to introduce additional constraints that may limit the number of spatial dimensions to three.

Some may note that the current approach fixes the signature of the Minkowski metric to be $(+,-,-,-)$ rather than the $(-,+,+,+)$ more commonly employed in space-time physics.  If treated simply as a metric, the signature is arbitrary.  However, here the signature arises from the symmetry and antisymmetry of the chain-like and antichain-like components of the interval pair.  The fact that these quantities have deeper significance does away with any arbitrariness of the resulting signature.  Moreover, the signature derived here agrees with the signature associated with the decomposition of mass into energy and momentum in particle physics.  This is significant \red as additional results suggest \black that these quantities \red are \black analogous to rates of interactions in the poset picture \cite{Knuth:FQXI2013}\cite{Knuth:Info-Based:2014}.

Another important insight is the fact that the quantity $\beta$, which is the poset analogue of speed, is the relevant quantity to describe two linearly-related pairs of chains.  In physics, speed has been demonstrated experimentally to be a relevant quantity.  Here it is recovered theoretically from the pair transformation and the identification of symmetric pairs with time and antisymmetric pairs with space. It is from this identification of $\beta$ with speed that we are led to the realization that there exists a finite maximum speed given by $\beta = \pm 1$.  This is exemplified by projection-like intervals where the two events defining the interval endpoints project to the same event on a quantifying chain.  In addition, the linearity of the pair transformation ensures that the length of the projected interval is always zero so that an interval representing motion at the maximum speed in one frame represents motion at the maximum speed in all linearly-related frames. \red As a result, the constant speed of light is derived rather than taken as given. \black

More importantly, in the poset picture one views motion in terms of connectivity, which implies that motion is a manifestation of interaction.  Particles don't move in this picture.  Instead, they transition via discrete jumps with every interaction \cite{Knuth:FQXI2013}\cite{Knuth:Info-Based:2014}.  This is a bit different from the usual conception of interaction as force.  It suggests that the interactions we have quantified, which appear to generate an emergent space-time, do not necessarily cause forces, but at least in some cases enable motion.  Due to the intimate connection with space-time, one is led to hypothesize that these interactions may represent a fundamental process from which gravity arises as some sort of side-effect, and as such one might be able to understand the equivalence between inertial and gravitational masses.  Such an idea would not be without precedent since it has been suggested that crystal dislocations result in gravity-like behavior in the analogous space-time experienced by an electron moving in graphene \cite{Mesaros-graphene-gravity-2010}.  It is possible that the theory presented here may have something to say about analogous space-time in special materials, such as graphene, by considering interactions between the electrons and the crystal lattice as defining a local space-time.

The metric derived here is the Minkowski metric which represents flat space-time.  It is important to note that the critical \red conditions in its derivation are \black the concepts of coordinated chains and linearly-related chains.  One does not need to have a universe in which there are chains that enjoy these relationships.  Instead our consistency requirement states that \emph{if} there are such chains, \emph{then} they should agree on the scalar measure of the set of mutually quantifiable intervals.  It could very well be that there are posets where strict coordination is not possible.  Additional interactions that act differentially on the coordinated chains will undoubtedly disrupt the coordination condition thus limiting the property of flatness to shorter finite ranges along the chains.  We hypothesize that such disruption may be interpreted in terms of a discrete version of curvature.

The discrete nature of interactions in this picture is reminiscent of quantum mechanics (certainly quantum electrodynamics).  It would be interesting if the poset picture, in which space-time emerges, could equally support quantum mechanics.  If this were the case, it may provide a foundation for quantum gravity.  We have demonstrated that a poset model of a free particle that influences its neighbors, but is not influenced by others, reproduces many aspects of Fermion physics \cite{Knuth:FQXI2013}\cite{Knuth:Info-Based:2014}.  Moreover, from this poset picture, one can derive the Feynman chessboard model of a particle~\cite{Feynman&Hibbs} that leads to the Dirac equation in 1+1 dimensions \cite{Knuth:FQXI2013}\cite{Knuth:Info-Based:2014}.  We are currently continuing these studies to determine how space-time physics is related to quantum mechanics in this poset picture.

Galileo wrote, ``Measure that which is measurable and make measurable that which is not so.''  This is the essence behind the concept of quantification.  Fundamental symmetries can place strong constraints on a quantification scheme, and in some cases the resulting constraint equations are analogous to what we call laws.  In this sense, such laws truly represent an underlying order.  By considering a minimalist picture of a physical phenomenon that retains some subset of the fundamental symmetries, one can begin to explore which symmetries may be responsible for a given set of laws by deriving them through \red consistent quantification. \black  As such, the methodology applied here represents a \red novel \black approach to fundamental, or foundational, physics.

\section{Acknowledgements}
Kevin Knuth would like to thank Keith Earle, Ariel Caticha, Seth Chaiken, Adom Giffin, Philip Goyal, Jeffrey Jewell, Carlos Rodr\'{i}guez, Jeff Scargle, John Skilling and Michael Way for many insightful discussions and comments.
He would also like to thank Rockne, Ann and Emily Knuth for their
faith and support, Henry Knuth for suggesting that he `try using a J', and Lucy Knuth for decorating his notes.
Newshaw Bahreyni would like to thank Shahram Pourmand for his
helpful discussions and Mahshid Zahiri, Mohammad Bahreyni and
Shima Bahreyni for their continued support.  The authors would
also like to thank Giacomo Mauro D'Ariano, Alessandro Tosini,
Joshua Choinsky, Oleg Lunin, Margaret May, Patrick O'Keefe, Matthew Sarker, Cristi Stoica, and James Lyons Walsh for
valuable comments that have improved the quality of this work.

\bibliographystyle{amsplain}
\bibliography{C:/Users/KK952431/kevin/files/papers/bibliography/knuth}

\end{document}